\documentclass[twocolumn]{aastex6}
\usepackage{graphicx}
\usepackage{amssymb}
\usepackage{epstopdf}
\usepackage{textcomp,txfonts} 

\shortauthors{Pineda et al.}
\shorttitle{Brown Dwarf Aurorae}

\begin{document}
\title{A Panchromatic View of Brown Dwarf Aurorae}
\author{J. Sebastian Pineda \altaffilmark{1,2}, Gregg Hallinan\altaffilmark{2}, and Melodie M. Kao\altaffilmark{2} }

\affil{\altaffilmark{1} University of Colorado Boulder, Laboratory for Atmospheric and Space Physics, 3665 Discovery Drive, Boulder CO, 80303, USA}

\affil{\altaffilmark{2} California Institute of Technology, Department of Astronomy, 1200 E. California Ave, Pasadena CA, 91125, USA }

\begin{abstract}
Stellar coronal activity has been shown to persist into the low-mass star regime, down to late M-dwarf spectral types. However, there is now an accumulation of evidence suggesting that at the end of the main sequence there is a transition in the nature of the magnetic activity from chromospheric and coronal to planet-like and auroral, from local impulsive heating via flares and MHD wave dissipation to energy dissipation from strong large-scale magnetospheric current systems. We examine this transition and the prevalence of auroral activity in brown dwarfs through a compilation of multi-wavelength surveys of magnetic activity, including radio, X-ray, and optical. We compile the results of those surveys and place their conclusions in the context of auroral emission as the consequence of large-scale magnetospheric current systems that accelerate energetic electron beams and drive the particles to impact the cool atmospheric gas. We explore the different manifestation of auroral phenomena in brown dwarf atmospheres, like H$\alpha$, and define their distinguishing characteristics. We conclude that large amplitude photometric variability in the near infrared is most likely a consequence of clouds in brown dwarf atmospheres, but that auroral activity may be responsible for long-lived stable surface features. We report a connection between auroral H$\alpha$ emission and quiescent radio emission in ECMI pulsing brown dwarfs, suggesting a potential underlying physical connection between the quiescent and auroral emissions. We also discuss the electrodynamic engines powering brown dwarf aurorae and the possible role of satellites around these systems to both power the aurorae and seed the magnetosphere with plasma.
\end{abstract}

\section{Introduction}

Within the past 15 years, the discovery and follow up observations of radio emission from brown dwarfs \citep[e.g.,][]{Berger2001, Hallinan2007, Route2012, Hallinan2015, Kao2016} have heralded a shift in our understanding of magnetic activity in low-mass stars and ultracool dwarfs (UCDs; spectral type $\ge$ M7). The accumulating evidence now suggest that there may be a transition at the end of the main sequence away from coronal/chromospheric Solar-like magnetic activity towards auroral planet-like phenomena, from current systems driven by local photospheric plasma motions to those driven by a global electrodynamic interaction in the large-scale magnetosphere. Moreover, the progress of several observational surveys of brown dwarfs across the electromagnetic spectrum allows us to put together a comprehensive view of UCD auroral phenomena, for the first time, in this article. However, in order to put these observations in context, it is essential to discuss both the standard coronal/chromospheric picture of stellar magnetic activity and the underlying processes that govern aurorae in the gas giant planets of the Solar System.

\subsection{Stellar Activity in Low-Mass Stars}\label{sec:stellar}

Our understanding of stellar magnetic activity is rooted in our understanding of the Sun. Solar observations of a host of phenomena, from impulsive flare events to long-term monitoring of sunspots as well as the study of coronal and chromospheric structures have formed the basis for interpreting observations of similar activity in M dwarfs \citep[e.g.,][]{Haisch1991}. Observations indicate that a version of the same mechanisms powering Solar magnetic activity operates in low-mass stars. The process requires an internal dynamo that generates the persistent magnetic field anchored deep in the stellar interior and non-thermal local heating of the upper atmosphere, above the photosphere, through magnetic reconnection and/or MHD wave dissipation \citep[e.g.,][]{Linsky1980}. 

In early M-dwarfs, with partially convective interiors, the dynamo is thought to be the same as operates in the Sun, the $\alpha \Omega$ dynamo, which depends in part on the shearing layer between the radiative core and the convective envelope to transfer rotational energy into magnetic energy, linking the magnetic activity to the rotation and internal structure of the star \citep[e.g.,][]{Ossendrijver2003, Browning2006}. This leads to a strong feedback between a star's rotational evolution, due to angular momentum loss in a stellar wind, and observable tracers of magnetic phenomena \citep[e.g.,][]{Covey2011, Reiners2012b}. For example, younger and more rapidly rotating M-dwarfs flare more frequently than similar older stars, depositing energy in their upper atmospheres at a higher rate early in their lifetimes \citep{Hilton2010}. The connection is further observed as a strong correlation between stellar rotation/age and emission lines that trace upper atmospheric heating \citep[e.g.,][]{Skumanich1972}. Emission features such as Ca \textsc{ii} H and K, and H$\alpha$ are more prevalent and stronger in faster rotating M-dwarfs as compared to slower rotators \citep{Delfosse1998, Mohanty2003, West2008, Browning2010,West2015}. 

Although these features for early M-dwarfs, are indicative of chromospheric atmospheric structures, their decline in slowly rotating stars does not indicate the disappearance of inverted atmospheric temperature profiles. As indicated by observations at ultraviolet (UV) wavelengths, chromospheric, transition region, and coronal emission lines, such as Mg \textsc{ii}, at 2796 \AA, N \textsc{v}, at 1239 \AA~and 1243 \AA, and Fe \textsc{xii}, at 1242 \AA, are prevalent in M-dwarf atmospheres, even for slowly rotating M-dwarfs that are `inactive' in H$\alpha$ \citep{France2013, France2016}. In some M-dwarfs, H$\alpha$ in absorption may actually reflect weak chromospheric activity \citep{Cram1985}. Indeed, many weakly active M-dwarfs with Ca \textsc{ii} H and K emission lines are known to display H$\alpha$ absorption features \citep[e.g.,][]{Walkowicz2009}. The presence of coronal structures, like those of the Sun, in M-dwarf atmospheres, is further corroborated by the detections of X-ray emission in observations of early M-dwarfs \citep[e.g.,][]{James2000,Pizzolato2003}. Like the the optical emission features, UV and X-ray emission is also strongly correlated with rotation/age, with observations showing constant emission levels for young objects rotating more quickly than $\sim$$5$ d, and the emission declining for more slowly rotating objects as they age \citep{Pizzolato2003, Shkolnik2014a, Cook2014}. 

Early M-dwarf radio emission also appears to be consistent with this coronal/chromospheric picture. From F-type dwarf stars to early M-type dwarf stars, the G\"{u}del-Benz relation demonstrates a tight empirical relation between coronal X-ray and quiescent radio emission, illustrating a deep connection between the coronal plasma producing the X-ray emission and the non-thermal energetic electrons responsible for the radio emission \citep{Gudel1993}. The persistent heating of this coronal plasma to over 10$^{6}$ K is typically associated with strong small-scale fields and their turbulent reconnection \citep[e.g.,][]{Rosner1985, Solanki2006}. Indeed, the results of Zeeman Doppler Imaging (ZDI) studies demonstrate that active flaring early M-dwarfs exhibit complicated non-axisymmetric multi-polar large-scale fields similar to what is seen on the Sun, suggesting the presence of significant magnetic structures that heat and power the coronal radio and X-ray emission \citep[e.g.,][]{Donati2008}.

\subsection{Auroral Processes in Planetary Magnetospheres}\label{sec:auroproc}

In contrast to the stellar paradigm, planetary auroral emissions are associated with large-scale field-aligned current systems that pervade the extended magnetosphere, connecting the planetary atmosphere to energetic processes in the middle magnetosphere. In the Solar System, there are three main mechanisms that generate auroral currents \citep[see][and references therein]{Keiling2012}. Firstly, the interaction between the solar wind and a planetary magnetosphere triggers magnetic reconnection events that accelerate electrons along the magnetic field lines. This mechanism dominates the aurorae of the Earth and Saturn \citep[e.g,][]{Cowley2004}. Secondly, the relative motion of an orbiting satellite through a planet's magnetosphere creates a current system in the flux tube connecting the moon and the planet. This mechanism produces the auroral emission associated with the moons Io and Enceladus of Jupiter and Saturn, respectively \citep[e.g.,][]{Saur2004}. Lastly, the breakdown of co-rotation between a rotating plasma disk and the planetary magnetosphere can create a shearing layer that drives auroral currents. This is the mechanism that powers the main Jovian auroral oval \citep[e.g.,][]{Cowley2001}. Moreover, the different mechanisms can overlap, as they do in the Jovian magnetosphere. Each of these electrodynamic engines generates strong field-aligned currents that drive accelerated electron beams, the fundamental ingredient of auroral emission processes. 

The acceleration of the electrons creates an energetic non-thermal energy distribution and can lead to the onset of the electron cyclotron maser instability (ECMI). The necessary criteria are an energy distribution dominated by the non-thermal component and a cyclotron frequency larger than than the local plasma frequency, 

\begin{equation}
\frac{\omega^{2}_{pe}}{\omega^{2}_{ce}} = \frac{4\pi n_{e} m_{e} c^2}{B^2} \le 1 \; ,
\label{eq:freq}
\end{equation}

\noindent where $n_e$ is the electron density, $B$ is the magnetic field strength, $m_{e}$ is the electron mass and $c$ is the speed of light \citep[see ][]{Treumann2006}. As the ratio of Equation~\ref{eq:freq} approaches unity, the maser becomes weaker and less efficient. However, under the conditions of a dilute plasma immersed in a strong magnetic field, the energetic electrons become an efficient radiation source. The result is a strong coherent radio source emitting near the local cyclotron frequency that is highly circularly polarized and beamed into a thin ($\sim$$1^{\circ}$) conical sheet with large opening angles, nearly perpendicular to the magnetic field direction, $\gtrsim$$80^{\circ}$ \citep{Dulk1985, Treumann2006}. ECM radio emission has been observed in the magnetized planets of the Solar System, signaling the presence of non-thermal energetic electron distributions in the regions around the planetary magnetic poles, near the top of the atmosphere \citep[e.g.,][]{Zarka1998}. 

The energetic electron beams responsible for the radio emission precipitate into the atmosphere and generate a cascade of additional auroral emission processes \citep[see][and references therein]{Badman2015}. In Jupiter and Saturn, where the atmospheres are predominantly hydrogen, the collision of the energetic electrons with the atmospheric gas leads to excitation and ionization of H/H$_{2}$ and subsequent emissions at UV and optical wavelengths, including Lyman and Balmer line emissions \citep{Perry1999, Vasavada1999, Grodent2003, Gustin2013, Dyudina2016}. The creation of ionized species in the Jovian and Kronian auroral regions also leads to significant ion chemistry within the atmosphere and the creation of the strongly emitting species H$_{3}^{+}$ \citep[e.g.,][]{Perry1999}. In Jupiter the ro-vibrational transitions of H$_{3}^{+}$ serve to effectively cool the atmosphere and regulate exospheric temperatures \citep{Maillard2011}. The deposition of energy from the electron beam into the atmosphere also leads to a significant thermal contribution to the auroral emissions between 7 $\mu$m and 14 $\mu$m \citep{Bhardwaj2000}. X-ray emission has also been detected in the auroral polar regions, a consequence of charge exchange reactions of highly ionized species such as oxygen and sulfur, likely created during ion precipitation in auroral currents \citep[e.g.,][]{Gladstone2002, Hui2009}. 

These different multi-wavelength auroral emission processes are the consequence of the energy dissipation from the electrodynamic engine operating in the planetary magnetosphere. In the Jovian system the bulk of the energy, $\sim$85\% goes into atmospheric heating and thermal radiation \citep{Bhardwaj2000}. Most of the remaining $\sim$15\% emerges as part of the UV emission, with less than $\sim$1\% of the energy going into the optical aurorae \citep{Bhardwaj2000}. Additionally, the radio contribution only represents $\lesssim$0.1\% of the total auroral energy, and the X-rays even less \citep{Bhardwaj2000}.

\subsection{Brown Dwarfs: Between Stars and Planets?}

The divide between stars and planets reflects the different natures of the atmospheres and physical properties of these objects. However, brown dwarfs, as objects which span this separation, constitute a regime in which there could be a transition from the planetary regime to the stellar one. Observationally, many of the the atmospheric properties of brown dwarfs, such as effective temperature, $T_{\mathit{eff}}$, overlap with those of, on the low-mass end, gas giant planets, and, on the high-mass end, very-low mass stars \citep[e.g.,][]{Burrows2001}. Since brown dwarfs cool over time, with core temperatures insufficiently high for sustained hydrogen burning throughout their lifetimes, individual objects may display $T_{\mathit{eff}}\sim2700$ K at early ages, but much cooler $T_{\mathit{eff}}\sim1000$ K at later ages, depending on the brown dwarf mass \citep[][]{Burrows2001}. This property makes it difficult to distinguish individual objects without mass/age measurements, and consequently a field population of brown dwarfs may be composed of a mix of objects with different ages and masses despite having similar effective temperatures \citep[e.g.,][]{Burrows2001}.

While it is possible that a distinct form of magnetic phenomena is manifest in the brown dwarf regime, the similarities in atmospheric properties makes it plausible that magnetic phenomena may also change continually across the brown dwarf regime from planets to stars. Indeed, the underlying magnetic dynamos of giant planets, brown dwarfs and very-low mass stars may be very similar \citep{Christensen2009, Morin2011}; however this idea is currently being tested \citep[see][]{Kao2016}. Nevertheless, the nature of the transition in magnetic activity across the brown dwarf regime is an open question, as is its dependence on physical properties such as mass and age. With the discovery of cooler and lower-mass brown dwarfs, and new evidence pointing to a breakdown of the coronal/chromospheric solar-like paradigm of magnetic activity, we are further motivated to consider the activity of brown dwarfs from the auroral-planet perspective. Consequently, both stellar and planet perspectives can be used to elucidate the nature of brown dwarf magnetic processes. In Section~\ref{sec:trends}, we discuss how the multi-wavelength trends in magnetic activity shift in the UCD regime. In Section~\ref{sec:aurorae}, we examine the activity data in the context of auroral phenomena in brown dwarf atmospheres. Lastly, in Section~\ref{sec:conc}, we provide our conclusions, while summarizing our findings in Section~\ref{sec:summary}.

\section{Trends in UCD Magnetic Activity}\label{sec:trends}

The multi-wavelength features of stellar magnetic activity change at the end of the main sequence, for late M dwarfs and UCDs. The shift in observational features are a consequence of significant differences between the stellar and sub-stellar regimes, reflecting changes in the internal structure, the large-scale magnetic field topology, and the atmospheric fractional ionization.

\begin{figure}[tbp]
	\centering
	\includegraphics{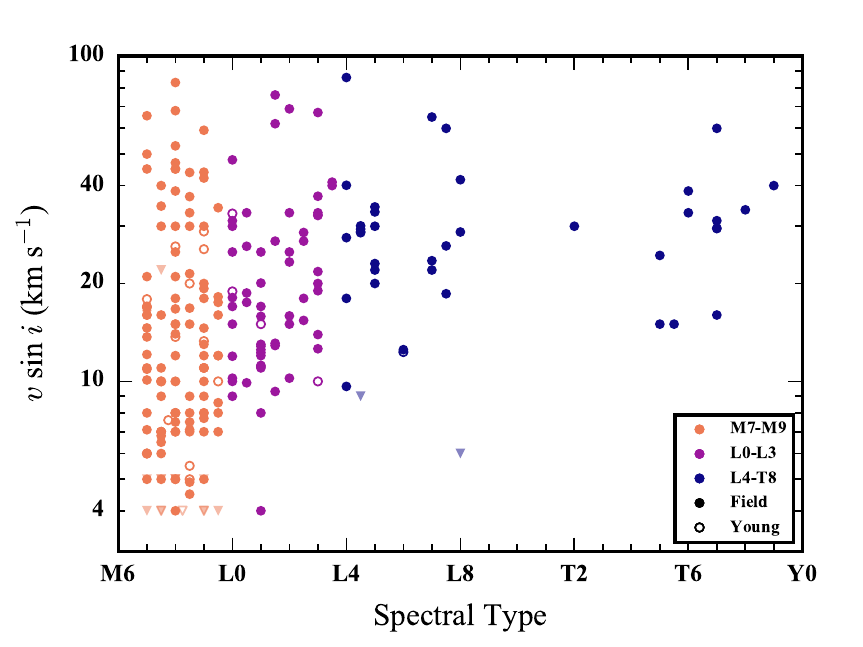} 
	\caption{The $\varv \sin i$ of UCDs as a function of spectral type, as compiled from the literature (see Appendix). The brown dwarfs all show fast rotation rates and short periods, even at typical field ages. Circles are detections, with filled versus open indicating field ages and indications of youth, respectively. Upper limits are plotted as triangles, with points grouped in different colors according to expected regimes of atmospheric ionization, see Section~\ref{sec:chromo}. }
	\label{fig:UCDVsin}
\end{figure}

\subsection{Convection, Dynamos, and Rotation}\label{sec:dynamo}

The lower mass and luminosity of very low-mass stars and brown dwarfs, relative to earlier type stars, have significant consequences for their internal structure. In contrast to early-M dwarfs, these objects have convective interiors extending from their cores through to their outer layers. Consequently, a distinct dynamo mechanism must operate in this fully convective regime (mass $\lesssim$0.3 M$_{\odot}$, spectral type $\gtrsim$dM4; \citealt{Chabrier2000}), in order to sustain the observed kG magnetic field strengths of these objects \citep{Reiners2007}. One commonly invoked dynamo is the $\alpha^{2}$ dynamo, which harnesses convective motions and rotation, but models have identified alternate dynamo mechanisms depending on a range of properties, including rotation rate and bolometric luminosity \citep{Browning2008, Christensen2009, Yadav2015}. Interestingly, despite the transition in the internal structure, there does not appear to be an abrupt change in the strength of the magnetic activity emission indicators across the fully convective boundary \citep[e.g.,][]{West2015,Wright2016,Newton2017}.

This transition coincides with a change in the prevalent magnetic field topologies. ZDI observations of fully convective M-dwarfs show the emergence of strong large-scale, dipolar fields in mid M-dwarfs as compared to the multi-polar fields of earlier stars \citep{Donati2006,Morin2010}. This appears to persist into the UCD regime, where both kinds of field topologies have been observed, suggesting either a bi-stability of the dynamo mechanisms or potential phase transitions between dynamo modes \citep{Morin2011,Kitchatinov2014}. However, current ZDI observations are only available for objects of M9 spectral type or earlier, often limited by the faint luminosities and fast rotation rates of many brown dwarfs, requiring an extrapolation of the likely field topologies to cooler objects \citep[see][]{Kao2016}. 

This change in topology had been thought to potentially drive changes in the angular momentum evolution of mid-to-late M-dwarfs as seen in the observed distribution of rotation periods, $P$, and projected rotational velocities, $\varv \sin i$ \citep{Irwin2009}. However, \citet{Reiners2012b} suggested that the rise in rotation rates of fully convective stars as compared to earlier stars may be driven predominantly by changes in the stellar radius. The increase in the observed rotation rates of very-low mass stars extends throughout the UCD regime where the object radius is nearly independent of mass, and is similar to the radius of Jupiter \citep[e.g.][]{Chabrier2000}. In Figure~\ref{fig:UCDVsin}, we plot a compilation of numerous literature sources (see Appendix) of $\varv \sin i$ measurements, illustrating the large rotational velocities of UCDs, and correspondingly their fast rotation rates ($\varv \sim~20$ km s$^{-1}$~$\rightarrow P \sim 6.2$~hr, for a radius of 1 $R_{\mathrm{Jup}}$). Indeed, most UCDs show signs of short rotation periods, with many showing periods of only a couple of hours, sustaining these fast rotation rates even at field ages \citep{Hallinan2008, Metchev2015}. This further suggests that UCDs do not have strong stellar winds that remove angular momentum as they do in stars, likely a consequence of the largely neutral atmospheres and diminished coronal activity (See Section~\ref{sec:chromo}). Traditionally, the Rossby number, $Ro = P/ \tau_{c}$, where $\tau_{c}$ is the convective overturn timescale, is used to quantify the effect of rotation on magnetic activity in stars; however, following \citet{Kiraga2007} and \citet{Reiners2010} the convective overturn timescale is a constant in the UCD regime and may not even be well defined for these objects. Consequently, we use the rotational velocity, as a broadly available observable, to compare the effects of rotation on magnetic phenomena amongst UCDs.

\subsection{Chromospheres and Coronae?}\label{sec:chromo}

The lower luminosities of very-low mass stars and brown dwarfs leads to much cooler $T_{\mathit{eff}}$, and correspondingly, much less ionized atmospheres. \citet{Mohanty2002} used atmospheric models to show that atmospheres below 2300 K are insufficiently ionized to support atmospheric current systems that sustain chromospheric and coronal activity. Recently, \citet{RodriguezBarrera2015}, taking a similar approach with more recent atmospheric models, suggested that the corresponding threshold is closer to a temperature of 1400 K, with variations depending on the atmospheric gravity and metallicity. Moreover, they show that, in these objects, significant portions of the atmosphere can be dominated by electromagnetic interactions, since the typical cyclotron frequency (assuming kG field strengths) far exceeds the collision frequency of electrons with neutrals \citep{RodriguezBarrera2015}.

These theoretical studies have important implications for the observed magnetic emissions of UCDs. We thus use these temperature thresholds as guides when considering the changes in the observed properties of UCDs with spectral type ($T_{\mathit{eff}}$), grouping the M-dwarfs (M7-M9), early L dwarfs (L0-L3) and late L dwarfs and T dwarfs (L4-T8) together (e.g.~Figure~\ref{fig:UCDVsin}). Although, an effective temperature of 2300~K roughly coincides with the M/L transition, we base our groupings in spectral type additionally on the observational findings of \cite{Pineda2016a} and \cite{MilesPaez2017} with regard to the prevalence of H$\alpha$ emission across the UCD regime, which suggest the break below which ionization is too low to sustain chromospheric activity is closer to L4, $T_{\mathit{eff}}$ $\sim$ 1600~K. We focus on H$\alpha$ and radio emission in Sections~\ref{sec:mag_Ha} and~\ref{sec:star_rad}, respectively, as the most extensively studied tracers of magnetic emissions in the UCD regime. 
	
	\begin{figure}[tbp]
		\centering
		\includegraphics{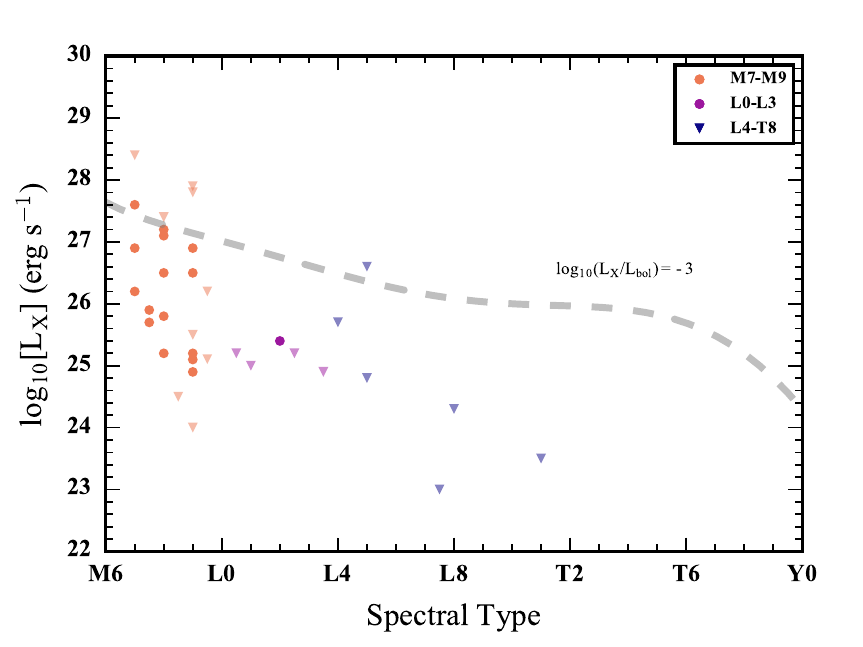} 
		\caption{The X-ray quiescent luminosities of UCDs as a function of spectral type as compiled from the literature (see Appendix). Objects with X-ray detections typically also show flaring emission, see \citet{Williams2014}. Relative to earlier-type stars, there is a steep drop off in observed X-ray emission for UCDs, with only one detection in objects later than L0. The dashed line indicates the typical X-ray emission level of active early M-dwarfs \citep{Berger2010}. }
		\label{fig:UCDXray}
	\end{figure}

However, the changes in the ability of UCDs to sustain significant magnetic heating of their upper atmospheres relative to stars is also manifest in additional multi-wavelength observations of magnetic phenomena. Photometric UV data on late M-dwarfs reveal predominantly weak NUV and FUV emissions \citep{Jones2016}, with spectroscopic data on an few targets showing transition region emission features in the FUV \citep[][]{Hawley2003}. The UV emissions of even cooler objects, L dwarfs and later, have remained largely unexplored. The UCD regime also exhibits a steep drop in X-ray emission relative to earlier-type stars \citep[e.g.,][]{Berger2010,McLean2012, Williams2014}. In Figure~\ref{fig:UCDXray}, we plot the quiescent X-ray luminosity of UCDs as a function of spectral type as compiled in the literature (see Appendix). Although some individual objects at the end of the main sequence are capable of heating a high temperature corona, the overall lack of X-ray and UV detections, despite the fast rotation rates of the objects in these samples, points towards the diminishing ability of UCDs to sustain hot coronae. 

Additionally, observations of flare events in UCD atmospheres suggest that the same processes generating flares in M stars continues to operate on some UCDs. These flare events are evident in the red optical data, noting the large increase in Balmer line emissions during short time intervals \citep[e.g.,][]{Liebert1999,Schmidt2007}. Monitoring of L dwarfs with \textit{Kepler} and concurrent spectroscopic observations, revealed that these white light flares resemble the same kinds of events on earlier type M dwarfs \citep{Gizis2013, Gizis2016b}. The flares are just as energetic with energies as high a $\sim$10$^{32}$ erg and potentially as strong as $\sim$10$^{34}$ erg \citep{Gizis2013,Schmidt2014}. \citet{Gizis2013} also demonstrated that energetic flares occur less frequently on their target L dwarf by factors of $\sim$10-100 than what is observed on early M-dwarf flare stars. These observations may reflect the decreased ability of the cooler UCD atmospheres to build up and release energetic flaring magnetic loops from buoyant flux tubes that have risen from the deep interior \citep{Mohanty2002}. Although some brown dwarfs are still able to generate these flares, as with the X-ray coronae and UV transition region emission lines, this data would suggest even fewer such flaring UCDs amongst the cooler late L dwarfs and T dwarfs.

\begin{figure}[tbp]
	\centering
	\includegraphics{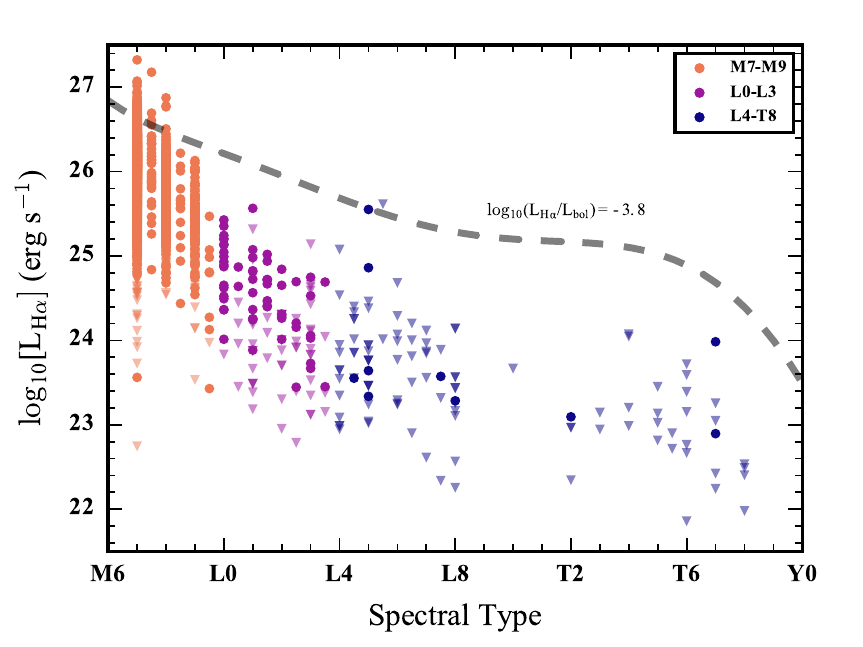} 
	\caption{The luminosity in H$\alpha$ in the UCD regime as a function of spectral type, spanning M7-T8, illustrating the decline in emission strength with effective temperature and the deviation from the typical stellar emission strengths. Detections are shown as filled circles and non-detections as triangles. We use the polynomial relations of \citet{Filippazzo2015} to determine the bolometric luminosity, $L_{\mathrm{bol}}$ as a function of spectral type. The different UCDs are further grouped in different colors according to expected regimes of atmospheric ionization, see Section~\ref{sec:chromo}. The data have been compiled from the literature (see Appendix).}
	\label{fig:UCDLHa}
\end{figure}

\subsubsection{H${\alpha}$}\label{sec:mag_Ha}

The H$\alpha$ emission of UCDs also diverges from that observed in stellar atmospheres. For `active' early M dwarfs, the strength of H$\alpha$ emission is roughly in line with a normalized level of $\log_{10} (L_{\mathrm{H}\alpha}/L_{\mathrm{bol}}) = -3.8$ \citep{Berger2010}; however, the strength of emission in cooler objects is much weaker and declines more rapidly than the bolometric luminosity. In Figure~\ref{fig:UCDLHa}, we show this decline by plotting the observed H$\alpha$ luminosity as a function of spectral type in the UCD regime with a dashed line indicating a constant level of $L_{\mathrm{H}\alpha}/L_{\mathrm{bol}}$ for early M dwarfs. The typical emission level departs considerably from the expected chromospheric value based on the bolometric luminosity. 

\begin{figure}[tbp]
	\centering
	\includegraphics{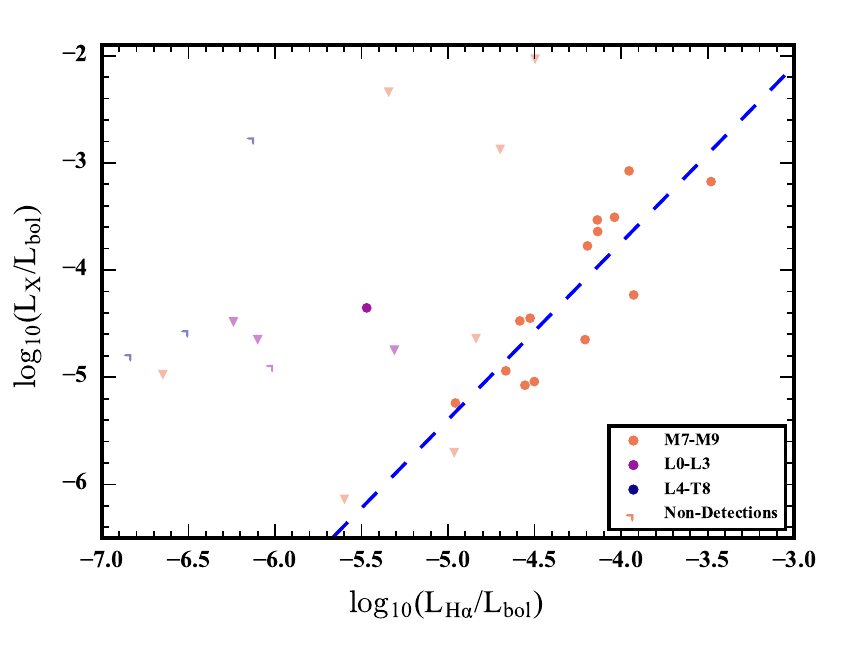} 
	\caption{The normalized X-ray luminosity as a function of H$\alpha$ luminosity in the UCD regime, showing all objects with measurements at both wavelength regimes in the literature (see Appendix). Amongst M7-M9 dwarfs there is a clear correlation between X-ray and H$\alpha$ emissions. The dashed line represents the best for line for these points of slope 1.65 and intercept of 2.86, with a scatter of 0.39 dex (see Equation~\ref{eq:xrayha}).}
	\label{fig:XrayLHa}
\end{figure}

This decline appears to be more gradual than what is observed in X-rays (see Figure~\ref{fig:UCDXray}), where the drop is more dramatic for L dwarf and cooler objects. In Figure~\ref{fig:XrayLHa}, we plot each object's X-ray emission against their H$\alpha$ emission, both normalized by their bolometric luminosities. For M7-M9 objects, there is a clear correlation between the observed H$\alpha$ and X-ray luminosities. The best fit line for these points is given as

\begin{equation}
\log_{10} (L_{X}/L_{\mathrm{bol}}) = 1.65 \, \log_{10} (L_{\mathrm{H}\alpha}/L_{\mathrm{bol}})  + 2.86 \; ,
\label{eq:xrayha}
\end{equation}

\noindent with a 0.39 dex scatter on the relation at fixed H$\alpha$ luminosity. Although we use typical values for these observed quantities, since the optical and X-ray observations are taken at different times, this scatter may be due predominantly to the intrinsic variability of the emission processes. Nevertheless the correlation shows a clear connection between coronal and chromospheric heating processes in the warmest UCDs, which does not appear to persist into the coolest objects, although more data are needed. If M7-M9 dwarfs behave like warmer stars, there is a limit to extrapolating this relation to high energies as the emission likely saturates, $\log_{10} (L_{X}/L_{\mathrm{bol}}) \sim$ -3 \citep[e.g.,][]{Pizzolato2003}.

\begin{figure}[tbp]
	\centering
	\includegraphics{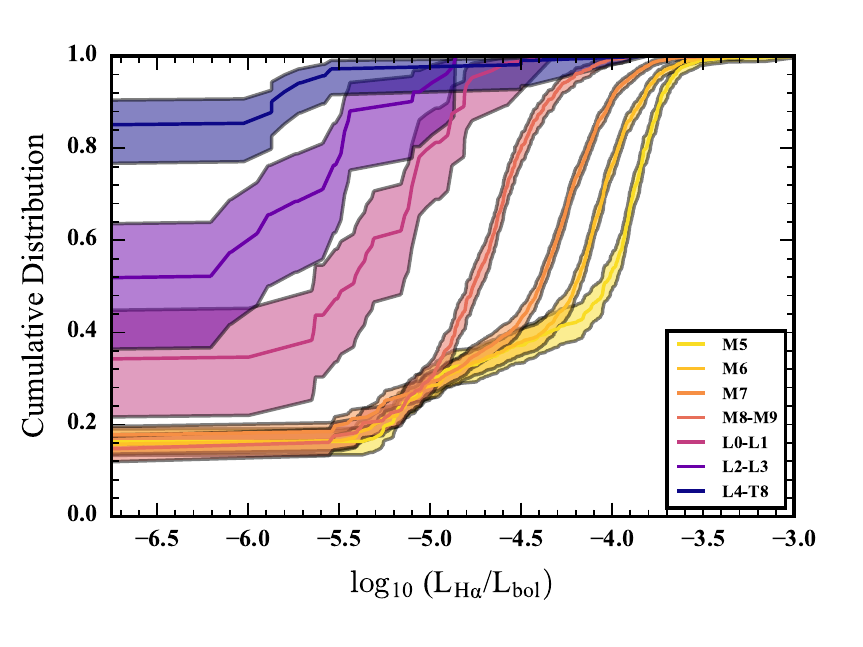} 
	\caption{Empirical cumulative distribution functions of H$\alpha$ luminosity normalized by the bolometric luminosity for late M-dwarfs and UCDs. Constructed using the Kaplan-Meier estimator, to account for non-detections, the curves illustrate the decline in the number of objects observed to be H$\alpha$ active and the declining strength of the emission (also see Figure~\ref{fig:UCDLHa}). The shaded regions represent 95\% confidence intervals. The shape of the distributions are also distinct going from low-mass stars to cool brown dwarfs, see Section~\ref{sec:mag_Ha}.}
	\label{fig:lumSeq}
\end{figure}

To better understand these trends in H$\alpha$ as a function of spectral type, we have constructed the empirical cumulative distribution functions (ECDFs) for this emission in UCDs and nearby mid M-dwarfs (from the Sloan Digital Sky Survey DR7 spectroscopic sample of \citealt{West2011}), using the Kaplan-Meier estimator, which takes into account non-detections. Equivalent width measurements consistent with 0 \AA~or indicative of absorption are treated as non-detections. In weakly active objects the emergence of a chromosphere initially manifests as stronger H$\alpha$ absorption before the line is filled in by stronger emission \citep{Cram1985}; although, for the coolest M-dwarfs the maximum observed absorption has equivalent width $\sim$0.075 \AA~\citep{Newton2017}. Consequently, in this analysis, some very weakly active objects are treated as non-detections. Although some observations may not have been sufficiently sensitive to detect very weak emissions, the majority of the data set consists of observations that probed deep enough to detect typical emission levels \citep[see][]{Schmidt2015,Pineda2016a}. Furthermore, the Kaplan-Meier estimator provides a robust statistical treatment of the non-detections that permits a comparison of the ECDFs. In Figure~\ref{fig:lumSeq}, we plot the ECDFs of $L_{\mathrm{H}\alpha}/L_{\mathrm{bol}}$ for objects with spectral types later than M4. The trend of cooler objects showing weaker emission is evident in how the curves shift to the left for later spectral type objects. Where each curve meets the ordinate axis indicates the fraction of H$\alpha$ non-detections in each spectral type bin. Clearly, the cooler objects are less frequently observed in emission. The M5-M9 objects show similar rates of H$\alpha$ detection, a consequence of restricting these bins to a height above the Galactic mid-plane, $|Z| <$100 pc, where the typical ages are below the activity lifetimes of these stars \citep{West2008}. When including more distant objects, the mid M-dwarfs show fewer `active' stars, and hence lower activity fractions \citep{West2008,Pineda2013b}. 

Because the ECDFs are constructed from all of the available literature, including many targets with only a single H$\alpha$ observation, the distributions include scatter associated with the intrinsic stochastic variability of UCD H$\alpha$ emission, as each individual object may have been in different activity states at the time of their respective observations. Consequently, the H$\alpha$ variability, which can change by factors of 1.2-4 on short timescales \citep{Lee2010}, does not dramatically impact our comparison of the ECDFs across the UCD regime, especially for the earlier-type objects with the larger sample sizes. The uncertainty in these distributions is captured by the shaded regions in Figure~\ref{fig:lumSeq}, and we thus treat these distributions as representative of their respective populations.

\begin{figure}[tbp]
	\centering
	\includegraphics{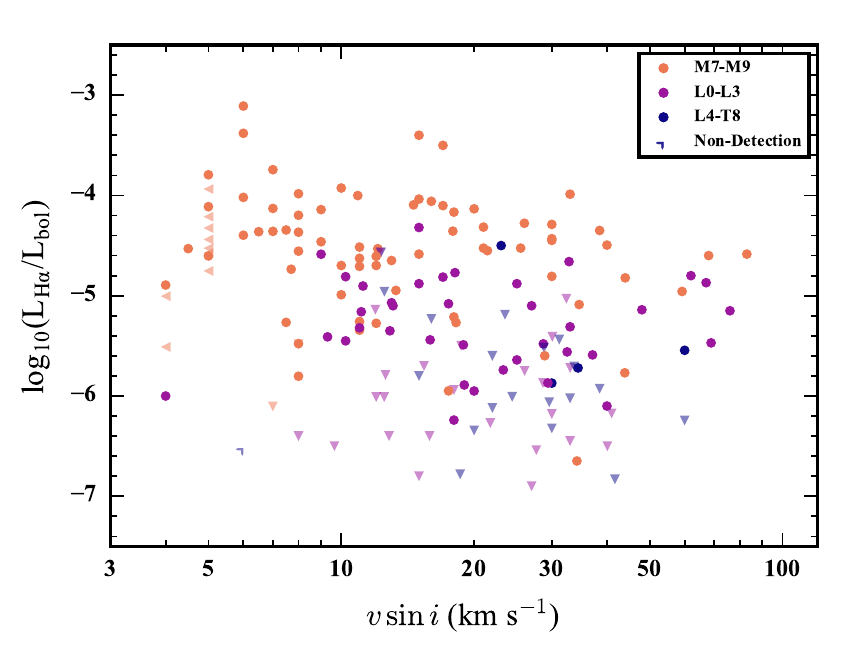} 
	\caption{The H$\alpha$ emission of UCDs normalized by their bolometric luminosity as a function of projected rotational velocity. For UCDs there does not appear to be a strong connection between H$\alpha$ emission and rotation. The data have been compiled from the literature (see Appendix). }
	\label{fig:UCDLHaVsin}
\end{figure}

In Figure~\ref{fig:lumSeq}, when comparing the cool brown dwarfs relative to the M dwarfs, there is a stark change in the shape of the H$\alpha$ distributions. While the shapes are similar amongst the M-dwarfs, the cooler brown dwarfs are observed with H$\alpha$ emission much less frequently, despite their rapid rotation (see Figure~\ref{fig:UCDVsin}). Moreover, observations explicitly show that in the UCD regime the H$\alpha$ emission is not strongly correlated with the rotation distribution \citep[e.g.,][and references therein]{Reiners2008,McLean2012}. In Figure~\ref{fig:UCDLHaVsin}, we show the distribution of UCDs with both H$\alpha$ measurements and $\varv \sin i$ measurements in the literature (see Appendix). There is no clear pattern of larger luminosities for faster rotators as has been observed in M dwarfs \citep[e.g.,][]{McLean2012,West2015}.

The change in the H$\alpha$ ECDFs for UCD is likely largely attributable to the effects of atmospheric ionization. The cooler objects magnetically heat their atmospheres less efficiently despite maintaining large rotational velocities, leading to distributions that show both weaker and less prevalent H$\alpha$ emission. \citet{Schmidt2015} further determined that this decline in activity coincided with a decreasing covering fraction for H$\alpha$ emitting regions in L dwarf atmospheres, which ostensibly vanish going into the T dwarf regime. However, this chromospheric picture does not explain the distribution for the coolest brown dwarfs in Figure~\ref{fig:lumSeq}. Despite having atmospheres which are too cool to sustain much ionization, there are still a few objects that have strong H$\alpha$ emission \citep{Burgasser2003, Pineda2016a}. This manifests in the ECDF for the L4-T8 bin extending to values of $\log_{10} (L_{\mathrm{H}\alpha}/L_{\mathrm{bol}})$ that are typical of late M-dwarfs (see Figure~\ref{fig:lumSeq}); consequently, the shape of the distribution for L4-T8 dwarfs does not fit neatly in the sequence defined by the predominantly chromospheric late M-dwarfs and early L-dwarfs. Although the difference in shape is driven by the strong emission of a few objects, like 2MASS~J1237+6526, the existence of these extreme outliers points towards potentially distinct behavior in L4-T8 dwarfs. Whereas the bolometric luminosity declines throughout the L4-T8 bin, unlike for earlier spectral types, the typical emission levels remain relatively constant across the L4-T8 range (see Figure~\ref{fig:UCDLHa}). These data indicate that the H$\alpha$ emission of the coolest brown dwarfs may be independent of $L_{\mathrm{bol}}$ and hence $T_{\mathit{eff}}$, unlike what is observed in stars with strong evidence for chromospheres/coronae.

These results suggest that a different mechanism is responsible for the H$\alpha$ activity in these coldest brown dwarfs, distinct from the chromospheric emission seen in stars. The activity distribution across the UCD regime can thus be characterized as being predominantly chromospheric for late M dwarfs, transitory across the early L dwarf sequence, and not chromospheric for late L dwarfs and T dwarfs.

\begin{figure}[btp]
   \centering
   \includegraphics{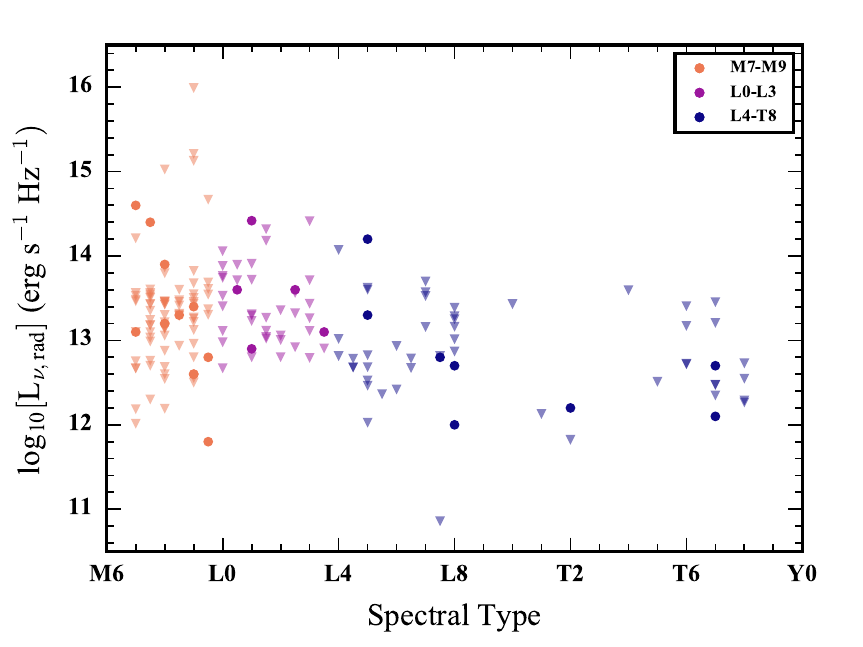} 
   \caption{The radio luminosity of UCDs as a function of spectral type, showing significant radio sources even for the coolest brown dwarfs. Triangles denote upper limits and filled circles correspond to detections. The plotted points only indicate the quiescent emission levels, with several objects also displaying strong radio bursts, see Tables~\ref{tab:ucds} and \ref{tab:ucds2}. The data are compiled from the literature (see Appendix). The different UCDs are grouped in different colors according to expected regimes of atmospheric ionization, see Section~\ref{sec:chromo}.}
   \label{fig:UCD_Rad}
\end{figure}

\subsubsection{Radio}\label{sec:star_rad}

\begin{figure}[tbp]
	\centering
	\includegraphics{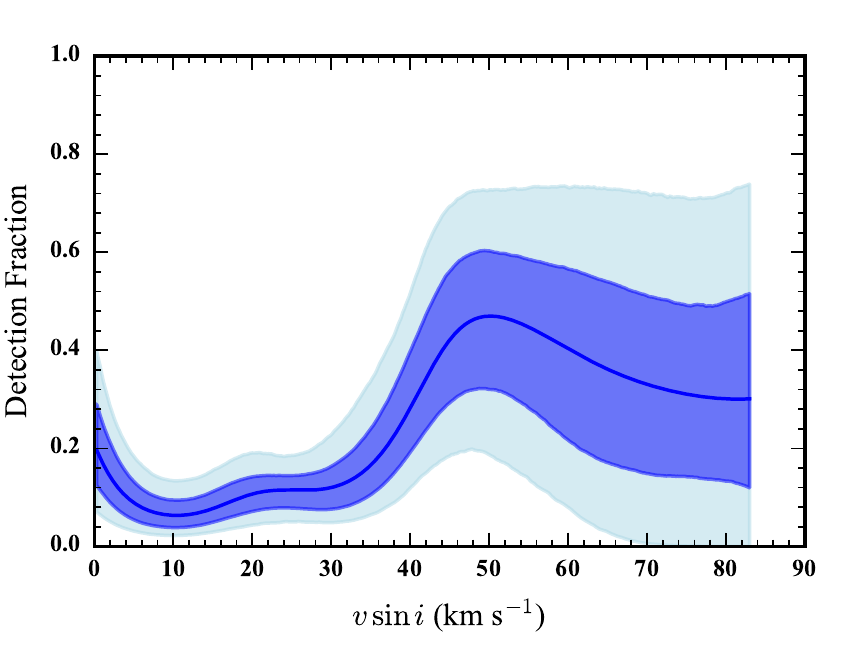} 
	\caption{The detection fraction of UCDs, M7-T8, in the radio as a function of $\varv \sin i$, illustrating a rise in radio detections for faster rotating objects. The fraction is computed by comparing the radio detections to the non-detections using a non-parametric adaptive kernel density estimation. The dark blue shaded region denotes the 68\% confidence interval, while the light blue region denotes the 95\% confidence interval using 5000 bootstrap samples of the set of targets with radio observations and $\varv \sin i$ measurements. }
	\label{fig:rad_frac}
\end{figure}

Unexpectedly, UCDs have also been observed to exhibit strong radio emission \citep{Berger2001}. Since the initial discovery, numerous surveys have looked for radio emission in very-low mass stars and brown dwarfs with very few detections \citep[e.g.,][]{Berger2006,McLean2012,Antonova2013}. In Figure~\ref{fig:UCD_Rad}, we plot a compilation of radio observations of UCDs in the literature (see Appendix) as a function of spectral type, showing only the quiescent radio luminosities in erg s$^{-1}$ Hz$^{-1}$, with observations typically conducted between 4 GHz and 9 GHz. As discussed in \citet{Route2016b}, there appears to be a general decline in the strength of the emission with spectral type, despite a large scatter in the observed luminosities. The overall detection rate of unbiased radio surveys is $\sim$7-10\% \citep{Route2016b}. In Figure~\ref{fig:rad_frac}, we plot the measured detection fraction as a function of $\varv \sin i$ for all UCDs with both radio observations and $\varv \sin i$ measurements\footnotemark[2]. This plot shows an increase in the observed number of radio emission detections for quickly rotating objects, expanding on the results illustrated by \citet{McLean2012}. While fast rotators can be disguised with slow rotational broadening due to high inclinations, the faster rotators could be rotating even more quickly. Interestingly, we see a sharp rise in the detection fraction at a $\varv \sin i$ of $\sim$40 km s$^{-1}$, which for objects with radii of 1 R$_{\mathrm{Jup}}$ and inclination close to 90$^{\circ}$, corresponds to a period of 3.1 hr. Although there may be an observational bias toward the detection of radio bursts (see below) in objects with rotation periods less than the typical radio observing duration limits (usually several hours), this does not necessarily preclude the detection of the quiescent radio emission (as plotted in Figure~\ref{fig:UCD_Rad}), consequently the trend evident in Figure~\ref{fig:rad_frac} represents a real rise in the radio detection rates for faster rotators.

\footnotetext[2]{We used the Adaptive Kernel Density Estimation routine \texttt{akj} within the \texttt{quantreg} package in R to construct probability density functions (PDF) for detections and non-detections as a function of $\varv \sin i$ and combined them to construct the detection fraction \citep{Rcore, Koenker2016}.}

\begin{figure}[tbp]
	\centering
	\includegraphics{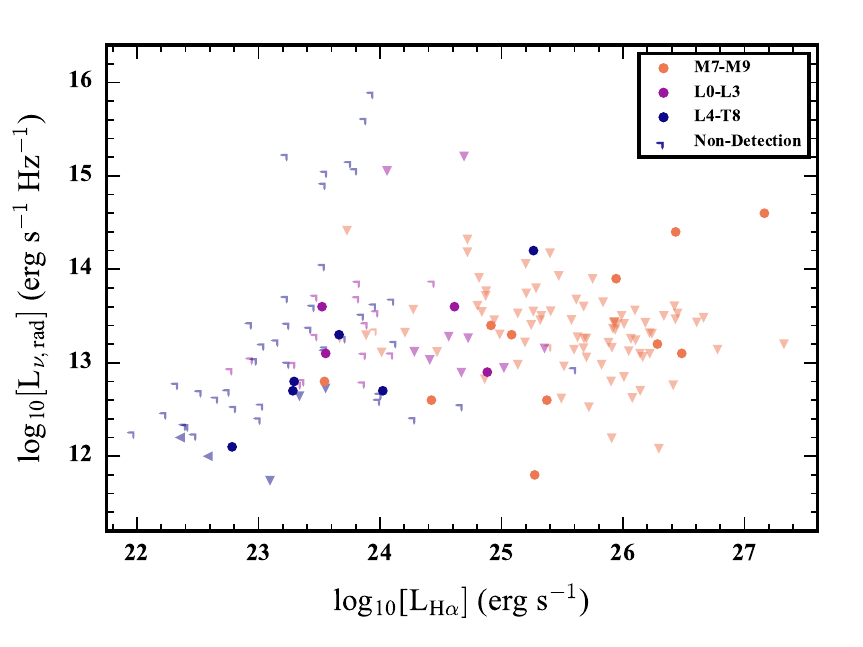} 
	\caption{The radio luminosity plotted against the H$\alpha$ luminosity in the UCD regime; no clear correlations evidenced for all of the data. Triangles denote upper limits and filled circles correspond to detections. Non-detections at both wavelengths are shown as corner symbols. The data are compiled from the literature (see Appendix), with the radio luminosity corresponding to quiescent emission levels as in Figure~\ref{fig:UCD_Rad}. The different UCDs are grouped in different colors according to expected regimes of atmospheric ionization, see Section~\ref{sec:chromo}}
	\label{fig:UCDHaRad}
\end{figure}

Significantly, the detected radio emission from these objects is morphologically distinct to that typically observed in low-mass stars. Quiescent stellar radio emission is likely dominated by a slowly varying gyro-synchrotron component at GHz frequencies, whereas stellar radio flares are highly energetic sporadic, impulsive events \citep[see][and references therein]{Gudel2002}. The observations of UCDs have shown short duration strong radio bursts at GHz frequencies super imposed on a quiescent background \citep[e.g.,][]{Route2012, Burgasser2015b}. However, studies that have observed individual UCDs, from M8.5-T6.5, for extended periods, have often discovered the bursts to be periodic, highly circularly polarized, and have high brightness temperatures, indicative of coherent emission \citep[e.g.,][]{Hallinan2007, Berger2009, Williams2015a}. Although the radio light curves show varying morphologies, the pulses are consistently periodic and in agreement with the ECMI. These radio emission properties provide direct evidence for the presence of stable auroral current systems, similar to those found in the Solar System giant planets (see Section~\ref{sec:auroproc}). We explore the aurorae and its consequences further in Section~\ref{sec:aurorae}. Although the acceleration mechanism is unclear, the quiescent radio components of UCDs appear consistent with gyro-synchrotron or synchrotron emission \citep{Williams2015b}.

The observations of pulsed ECMI radio emission throughout the UCD regime indicates the ability of these atmospheres to host conditions (see Equation~\ref{eq:freq}) amenable to the production of ECM radio sources, independent of the nature of UCD chromospheres and coronae (see below). The observed rate of periodically pulsing radio brown dwarfs is likely  influenced by the properties of the ECM radio beaming. Interestingly, objects that exhibit a quiescent radio component have also shown strong periodic radio pulses when monitored for times exceeding their rotational periods. For example, initial radio detections in short observations for TVLM513-46546, LSRJ 1835+3259, 2MASS~J0036+1821, 2MASS~J1047+2124, and NLTT 33370, were followed up in extended monitoring for several hours that found periodic radio pulses \citep{Hallinan2007, Hallinan2008, McLean2011, Route2012}. These results suggest that perhaps the pulses and the quiescent emission could be a consequence of the same underlying conditions. However, monitoring of other targets, such as DENIS~J1048.0-3956, LP~944-20, and 2MASS~J0952219-192431 did not detect periodic pulses, although this could be due to intrinsic variability, long rotational periods, and/or the effects of ECMI radio beaming \citep{McLean2012, Lynch2016}. 

\begin{figure}[tp]
	\centering
	\includegraphics{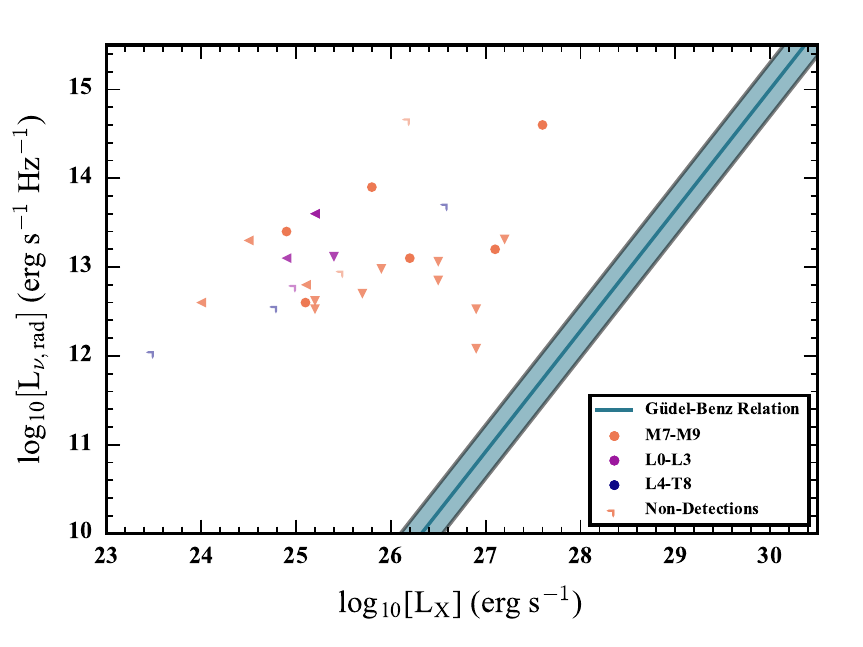} 
	\caption{The quiescent radio luminosity versus the quiescent X-ray luminosity of UCDs for objects observed at both wavebands. The shaded strip shows the G\"{u}del-Benz relation and scatter for the tight X-ray--radio correlation in stellar coronal activity \citep{Gudel1993,Williams2014}. The radio emitting UCDs strongly diverge from this empirical relation. }
	\label{fig:UCD_Gudel}
\end{figure}

The pulsed radio emission is distinct from typical stellar behavior and is related to the underlying mechanisms that generate the ECMI. Moreover, additional clues can be gathered from a comparison to other wavelength bands. We noted the low detection rate of radio UCDs; however, that rate is bolstered by the success of recent studies using a targeted sampled, based on H$\alpha$ emission in late L dwarfs and T dwarfs, suggesting a link between these emissions in the coldest brown dwarfs \citep{Kao2016}. We compare these two samples in Figure~\ref{fig:UCDHaRad}, showing the luminosity in H$\alpha$ versus the radio luminosity, as in Figures~\ref{fig:UCDLHa} and~\ref{fig:UCD_Rad}, respectively. For the majority of late M dwarfs, no radio emission is observed even when H$\alpha$ emission is present. Although these objects often show variable H$\alpha$ emission, the luminosities typically change by factors of only a couple \citep{Lee2010}. However, for L4-T8 objects, there is a significant overlap between objects showing both H$\alpha$ and radio emissions (also see Tables~\ref{tab:ucds} and~\ref{tab:ucds2}).

The changes in the UCD radio behavior are also evident when comparing these observations to the X-ray observations, both in quiescence, as shown in Figure~\ref{fig:UCD_Gudel}, which shows the G\"{u}del-Benz relation as a shaded stripe \citep{Gudel1993,Benz1994,Berger2010,Williams2014}. The low X-ray luminosities (see also Figure~\ref{fig:UCDXray}) would suggest only weak radio emissions, based on the coronal/chromospheric perspective; however, the observed quiescent radio luminosities are much stronger, indicating that these processes may no longer coupled in objects with these effective temperatures, $T_{\mathrm{eff}} \lesssim 2600$ K. Although these emissions may still be connected in some of the late M dwarfs, consistent with the portion of objects displaying H$\alpha$ chromospheric emission (see Section~\ref{sec:mag_Ha}, Figure~\ref{fig:XrayLHa}), the regime of magnetic activity as reflected by the radio emission is distinct. While the conditions capable of generating the ECMI and subsequent radio emissions are present, the conditions to generate substantial X-ray emission have vanished. Interestingly, some objects do show X-ray emission, and if it is attributable to weak coronal plasmas, the observed radio emission carries several orders of magnitude more energy than any associated synchrotron emission, as predicted by the G\"{u}del-Benz relation. Considering both the H$\alpha$-X-ray relation (see Figure~\ref{fig:XrayLHa}) and the the G\"{u}del-benz relation, there are several objects that are consistent with the former, but depart from the latter. If the radio is indeed decoupled from the X-ray for these objects, then these data could be explained by overlapping mechanisms in this regime, both coronal/chromospheric and ECMI related.

\subsection{Photometric Variability}\label{sec:photovar}

Traditional stellar photometric variability has been interpreted as evidence for star spots. However, as the atmosphere cools and becomes more neutral, the ability of the atmosphere to sustain these magnetic features becomes less clear. Concurrently, the low temperature of the atmospheres allows for the formation of dust condensates and clouds that influence the emergent stellar flux \citep[e.g.,][]{Marley2015}. Thus, in UCD atmospheres there is some ambiguity with regards to the dominant processes generating photometric variability, especially at the M/L transition, whether magnetic spots or clouds \citep{Gizis2015}. Doppler imaging of some example late M dwarfs by \citet{Barnes2015} revealed high latitude features which they interpreted as magnetic spots. However, these kinds of high latitude features, and the photometric variability they generate can also be interpreted as a consequence of an auroral electron beam (see Section~\ref{sec:bd_var}).

\begin{deluxetable*}{l c c  c c c c c p{0.1\linewidth} p{0.1\linewidth}}
	\tablecaption{ Ultracool Dwarfs with Rotationally Periodic Radio Pulses
		\label{tab:ucds} }
	\tablehead{
		\colhead{Object} & \colhead{SpT} & \colhead{T$_{\mathit{eff}}$} & \colhead{$[ L_{\mathrm{bol}}]$ } & \colhead{Mass } & \colhead{Period } & \colhead{$[L_{\nu}]$} & \colhead{$[L_{\mathrm{H}\alpha}/L_{\mathrm{bol}}]$\tablenotemark{c}}&   \colhead{Phot. Var.\tablenotemark{d}}  & \colhead{References\tablenotemark{e}} \\
		& & \colhead{(K)} & \colhead{ ($L_{\odot}$) } & \colhead{ (M$_{\mathrm{Jup}}$) }  & \colhead{(hr)\tablenotemark{a}} & \colhead{(erg s$^{-1}$ Hz$^{-1}$)\tablenotemark{b}}&&& }
	\startdata
	NLTT 33370\tablenotemark{f} & M7 & 2954 $\pm$ 3 & -2.616 &  92.8 $\pm$ 0.6& --- & $<12.89$  & $\sim$ -3.8 & MEarth, $g$, $i$ & 8, 8, \textbf{10}, \textbf{\underline{19}}, \\
	& M7 &  2947 $\pm$ 4 &  -2.631   &   91.7 $\pm$ 1.0 & 3.89  &  14.6, \textbf{(15.4)} &--- & --- & \textbf{\textit{\underline{26}}}  \\
	LSR~J1835+32\tablenotemark{g} & M8.5 & 2316 $\pm$ 51 & -3.50 & 77 $\pm$ 10 & 2.84& 13.3, \textbf{(14.0)} & $\sim$ -5.0 & $I$, $R$ & 23, 9, \textbf{1}, \textbf{12}, \underline{13}, \textit{24}  \\
	TVLM~513-46 &M8.5   & 2242 $\pm$ 55 &-3.57  &  75 $\pm$ 11  & 1.9596\tablenotemark{h} & 13.4, \textbf{(14.8)} &  $\sim$ -5.1  & $I$, $i'$, $z'$, $J$&  16, 9, \textit{5}, \underline{7}, \textbf{11}, \underline{13}, \textbf{20}, \textbf{27} \\
	J0746+2000\tablenotemark{gi} &L0.5  &  2205 $\pm$ 50 & -3.64 & 81.7 $\pm$ 4.2 & 3.32  & --- & $\sim$ -5.2 & $I$ & 3, 14, 18, \textbf{\textit{\underline{2}}},  \\
	& L1.5 &  2060 $\pm$ 70 & -3.77 & 76.5 $\pm$ 4.2  &  2.072 & 13.6, \textbf{(15.3)} & --- & --- &  \underline{13}, \textbf{18}\\
	J0036+1821\tablenotemark{g} & L3.5 & 1869 $\pm$ 64 & -3.93 & 66 $\pm$ 13 & 3.08 &  13.1, \textbf{(13.8)}  & $\sim$ -6.1& $R$, $I$, $z'$, $J$, $K_{s}$, 3.6, 4.5 & 17, 9, \underline{6}, \textbf{12}, \underline{13}, \textbf{20}, \underline{21}, \textit{22} \\
	J1047+2124\tablenotemark{g} & T6.5 & 880 $\pm$ 76& -5.30 & 42 $\pm$ 26 & 1.77 & 12.1, \textbf{(13.2)} & $\sim$ 5.5 & --- &  4, 9, \textbf{15}, \textit{22}, \textbf{25}
	\enddata
	\tablenotetext{a}{The periods are quoted from the radio data unless otherwise noted, and are consistent with the periods detected in photometric monitoring.}
	\tablenotetext{b}{Radio luminosity entries list the quiescent emission and in parentheses the typical pulse peak flux densities. There can be a wide degree of variability in the strength of the quiescent emission from epoch to epoch and in the strength/shape of the the radio pulses; the entries here give representative values for the average energy and peak luminosity of the quiescent emission and pulses, respectively.}
	\tablenotetext{c}{H$\alpha$ luminosities show variability in general; the entries listed here are representative values, usually taking the typical EW in the literature and converting to $L_{\mathrm{H}\alpha}/L_{\mathrm{bol}}$ using the $\chi$ values of \citet{Schmidt2014b}.}
	\tablenotetext{d}{This column lists the photometric passbands with confirmed variability; 3.6 and 4.5 refer to the \textit{Spitzer} IRAC bands at 3.6 and 4.5 $\mu$m. The MEarth broad passband is in the red optical \citep{Nutzman2008}.}
	\tablenotetext{e}{The first reference listed denotes the source for the spectral type, the next correspond to the physical properties, with the period and radio references in bold, the H$\alpha$ reference in italics and the photometric variability reference underlined.}
	\tablenotetext{f}{Physical properties are determined from models and the luminosities of the individual components \citep{Dupuy2016}. VLBI imaging confirms the radio emission to be from only the secondary with the corresponding pulse period \citep{Forbrich2016, Dupuy2016}. The H$\alpha$ emission and photometric variability cannot be attributed to particular components.}
	\tablenotetext{g}{The full names of these objects are, in table order, LSR~J1835+3259, 2MASSI~J0746425+200032, 2MASS~J00361617+1821104, and 2MASS~J10475385+2124234.}
	\tablenotetext{h}{Period precision is quoted by \citet{Wolszczan2014} as 7 ms.}
	\tablenotetext{i}{Physical properties combine dynamical measurements by \citet{Konopacky2010} and evolutionary models \citep{Harding2013b}. The period of the primary is determined from photometric monitoring, which distinguishes the radio emission to be from the secondary \citep{Harding2013b}. The H$\alpha$ emission is periodic at the same rotation period as the radio emission \citep{Berger2009}.}
	\tablenotetext{}{R\textsc{eferences.} -- (1) \citet{Berger2008}, (2) \citet{Berger2009}, (3) \citealt{Bouy2004}, (4) \citet{Burgasser2006}, (5) \citet{Burgasser2015a}, (6) \citet{Croll2016a}, (7) \citet{Croll2016b}, (8) \citet{Dupuy2016}, (9) \citet{Filippazzo2015}, (10) \citet{Forbrich2016}, (11) \citet{Hallinan2007}, (12) \citet{Hallinan2008}, (13) \citet{Harding2013a}, (14) \citet{Harding2013b}, (15) \citet{Kao2016}, (16) \citet{Kirkpatrick1995}, (17) \citet{Kirkpatrick2000}, (18) \citet{Konopacky2010}, (19) \citet{McLean2011}, (20) \citet{McLean2012}, (21) \citet{Metchev2015}, (22) \citet{Pineda2016a}, (23) \citet{Reid2003} , (24) \citet{Schmidt2007}, (25) \citet{Williams2015a}, (26) \citet{Williams2015c}, (27) \citet{Wolszczan2014}.  }
\end{deluxetable*}

Interestingly, photometric and spectroscopic monitoring of UCDs has revealed multi-wavelength broadband variability across the full range of spectral types from late M dwarfs to T dwarfs \citep[e.g.,][]{Harding2013a, Apai2013}. Many of these observations have taken place in near IR wavebands, predominantly J and K, with monitoring observations of several hours. The results of these studies have revealed several large amplitude variables located at the transition between L dwarfs and T dwarfs and some additional lower amplitude and less frequent variability in spectral types away from this transition \citep{Radigan2012, Radigan2014}. These observations have been interpreted as evidence for a patchy transition in cloud cover from L dwarfs to T dwarfs \citep[e.g.,][]{Artigau2009,Buenzli2014}. Additionally, \textit{Spitzer} monitoring of UCDs indicates that low-level variability is ubiquitous in the 3.6 and 4.5 $\mu$m bands, also interpreted as clouds \citep{Metchev2015}. Many of the IR variable objects also appear to display significant optical variability, with amplitudes of $\sim$10\%, comparable to the highest amplitude IR variables \citep{Biller2013,Heinze2015}

The connection between optical and infrared variability is potentially significant in the context of magnetic activity because brown dwarfs with confirmed ECM radio emission also appear to show very clear long-term photometric variability at optical wavelengths \citep{Harding2013a}. Moreover, \citet{Hallinan2015} demonstrated that the radio and optical variability may be linked as a consequence of auroral phenomena, distinct from standard starspot features (see Section~\ref{sec:bd_var}). 

Although it is possible that many of these processes are connected, both in optical and infrared, it is also probable that UCDs may display multiple sets of phenomena in different objects. One potentially distinguishing feature is the nature of the photometric variability. While much of the variability is seen to evolve with time, leading to irregular periodicity in extended monitoring, other observations show long lived steady structures \citep{Harding2013a, Crossfield2014n, Metchev2015, Gizis2015, Buenzli2015}. This difference may define a distinction between variability caused by magnetic spots and/or aurorae, and heterogenous cloud cover; however the relative importance of each across the UCD regime remains unclear. The possibility of long-lived storm systems in the cloudy atmospheres of brown dwarfs, adds an additional complication, potentially requiring concurrent multi-wavelength data to disentangle, such as H$\alpha$ emission in addition to the photometric monitoring \citep{Gizis2015}. Although, it is likely that clouds play a more important role for the coolest objects and that starspots may only be significant in the warmest UCDs, the nature of the transition from one to the other is an open question. Furthermore, as we will discuss in Section~\ref{sec:bd_var}, photometric variability as a consequence of aurorae may be present throughout the UCD regime.

\begin{deluxetable*}{l c c  c c c c  p{0.1\linewidth} p{0.1\linewidth}}
	\tablecaption{Ultracool Dwarfs with Quiescent Radio Emission or Unconfirmed Radio Pulse Periodicity
		\label{tab:ucds2} }
	\tablehead{ \colhead{Object} & \colhead{SpT} & \colhead{T$_{\mathit{eff}}$} & \colhead{$[ L_{\mathrm{bol}}]$ } & \colhead{Mass} & \colhead{$[L_{\nu}]$} & \colhead{$[L_{\mathrm{H}\alpha}/L_{\mathrm{bol}}]$\tablenotemark{b}}&   \colhead{Phot. Var.\tablenotemark{c}}  & \colhead{References\tablenotemark{d}} \\
		& & \colhead{(K)} & \colhead{ ($L_{\odot}$) } & \colhead{ (M$_{\mathrm{Jup}}$)}  & \colhead{(erg s$^{-1}$ Hz$^{-1}$)\tablenotemark{a}}&&& }
	\startdata
	J0952-1924 AB\tablenotemark{ef} & M7 & --- & --- & ---  &   14.4 & $\sim$ -4.0 & --- & 21, \textbf{35}, \textit{32} \\
	LHS 3003\tablenotemark{g} & M7 & 2595 $\pm$ 29 & -3.2 & ---  &  13.1 & $\sim$ -3.9 & --- & 4, 18, \textbf{5}, \textit{41} \\ 
	LHS 2397a\tablenotemark{g} &M8& 2461 $\pm$ 29 & -3.3 & ---  & 13.2 &$\sim$ -4.0& --- & 20, 18, \textit{39}, \textbf{43} \\
	& L7.5  & 1330 $\pm$ 29 & -4.5 & --- &--- & --- & --- & \\
	J1048-3956\tablenotemark{e} & M8 & 2307 $\pm$ 51  & -3.51 & 77.00$\pm$ 10.37  & 12.6, \textbf{(14.8)} & $\sim$ -4.7 & --- & 22, 18, \textbf{5}, \textbf{34}, \textit{10} \\  
	LP~349-25\tablenotemark{h}  & M8 &  2660 $\pm$ 30  & -3.04 & 67 $\pm$ 4 & 13.9 &$\sim$ -4.6& $I$, $R$ & 19, 16, \underline{27}, \textbf{35} \\ 
	& M9 & 2520 $\pm$ 30 & -3.18 & 59 $\pm$ 4& --- &  --- & --- & \\
	LP~944-20  & M9 & 1942 $\pm$ 144 & -3.56 & 29.53 $\pm$ 16.87  & 12.6, \textbf{(13.9)} & $\sim$ -5.6 & --- & 14, 18, \textit{10}, \textbf{34}, \textbf{35} \\
	J0024-0158\tablenotemark{e} & M9.5 & 2390 $\pm$ 80  & -3.44 & 79.27$\pm$ 11.13  & 12.8, \textbf{(13.8)} & $\sim$ -6.6 & $<I$ & 33, 18, \textbf{2}, \textit{10}, \underline{27} \\
	J0720-0846\tablenotemark{eg}& M9.5 & 2321 $\pm$ 113& -3.51 &  85 $\pm^{29}_{17}$  &11.8, \textbf{(12.9)}& $\sim$ -4.8 & $I+z$ & 11, 9, 18, \textbf{\textit{\underline{9}}}\\
	& T5.5 & 991 $\pm$ 113& -4.82 & 65 $\pm^{9}_{12}$ & --- &  --- & ---&  \\
	J1906+4011\tablenotemark{egi} & L1 &2102 $\pm$ 113 & -3.70 & --- & 12.9 & $\sim$ -5.0 & \textit{Kep}, $i$, $z$, 3.6, 4.5 & 23, 18, \textbf{\textit{24}}, \underline{25} \\ 
	J0523-1403\tablenotemark{e} &L2.5 & 1939 $\pm$ 68 & -3.86 & 67 $\pm$ 13  & 13.6 & $\sim$ -6.2 & $<I_{C}$ & 14, 18, \textbf{3}, \underline{31}, \textit{42}  \\ 
	GJ 1001 B\tablenotemark{g} & L5 & 1581 $\pm$ 113  & -4.22 & ---  & 13.3 & $\sim$ -5.7 & --- & 38, 18, \textbf{34}, \textit{42} \\ [2pt] 
	&  L5 & 1581 $\pm$ 113  & -4.22 & --- & --- & --- & --- &  \\
	J1315-2649\tablenotemark{eg} & L5& 1581 $\pm$ 113 &-4.22 & ---  &14.2 &$\sim$ -4.1 &$<J$,$<K'$ & 8, 18, \textbf{8}, \underline{29}, \textit{42} \\ 
	& T7 & 825 $\pm$ 113 & -5.1 & --- &--- &---&---& \\ 
	J0423-0414\tablenotemark{egj} & L6 & 1483 $\pm$ 113 & -4.39 & ---  & 12.8, \textbf{(13.7)} & $\sim$ -5.9 & $<I_{C}$, $J$, $K$ & 6, 18,  \underline{13}, \underline{17},  \\
	&  T2 & 1184 $\pm$ 113 & -4.62 & --- & --- & ---  & --- &  \textbf{28}, \textit{36}, \underline{44} \\ 
	J1043+2225\tablenotemark{egj}& L8 & 1336 $\pm$ 113 & -4.50 & ---  & 12.7, \textbf{(13.4)}& $\sim$ -5.8 & --- & 15, 18, \textbf{28}, \textit{36} \\ 
	J0607+2429\tablenotemark{eg}& L8 & 1336 $\pm$ 113 & -4.50  & ---  & 12.0 & $<-6.5$ & $<3.6$, $<4.5$, $<$ \textit{Kep.} & 12, 18, \textbf{\textit{\underline{26}}} \\ 
	SIMP 0136+0933\tablenotemark{ejk} & T2.5 & 1089$\pm^{62}_{54}$ & -4.63 & 23$\pm^{16}_{13}$ &  12.2, \textbf{(13.0)} &$<-6.6$& $J$, $H$, $K$ & 1, 28, \textbf{28}, \textit{36}, \underline{37} \\
	J1122+2550\tablenotemark{egjl} & T6 & 943 $\pm$ 113 & -4.9 & ---  & \textbf{(14.9)} & --- & --- & 30, 18, \textbf{40}  \\  
	J1237+6526\tablenotemark{ej} & T6.5 & 851 $\pm$ 74  & -5.36 & 40.85 $\pm$ 25.96  & 12.7, \textbf{(13.1)} & $\sim$ -4.2 & --- & 7, 18, \textbf{28}, \textit{36}
	\enddata
	\tablenotetext{a}{Radio luminosity entries list the quiescent emission and in parentheses the typical pulse peak flux densities. There can be a wide degree of variability, many of these objects have been observed multiple times, sometimes yielding non-detections; the entries here give representative values for the average energy and peak luminosity of the quiescent emission and pulses, respectively.}
	\tablenotetext{b}{H$\alpha$ luminosities show variability in general; the entries listed here are representative values, usually taking the typical EW in the literature and converting to $L_{\mathrm{H}\alpha}/L_{\mathrm{bol}}$ using the $\chi$ values of \citet{Schmidt2014b}.}
	\tablenotetext{c}{This column lists the photometric passbands with confirmed variability; 3.6 and 4.5 refer to the \textit{Spitzer} IRAC bands at 3.6 and 4.5 $\mu$m. \textit{Kep.} refers to the broadband optical passband of the \textit{Kepler} mission. Entries prefaced by `$<$' indicate constant fluxes during photometric monitoring observations.}
	\tablenotetext{d}{The first reference listed denotes the source for the spectral type, the next correspond to the physical properties, with the radio references in bold, the H$\alpha$ reference in italics and the photometric variability references underlined.}
	\tablenotetext{e}{The full names of these objects are, in table order, 2MASSW~J0952219-192431, DENIS~J1048.0-3956, 2MASSI~J0024246-015819, 2MASS~J07200325-0846499, 2MASS~J19064801+4011089, 2MASSI~J0523382-140302, 2MASS~J13153094-2649513, 2MASS~J04234858-0414035, 2MASS~J10430758+2225236, WISEP~J060738.65+242953.4, SIMP J013656.5+093347.3, WISEP~J112254.73+255021.5, and 2MASS~J12373919+6526148.}
	\tablenotetext{f}{This object is an unresolved spectroscopic binary, all measurements are from combined light \citep{Reid2002b,Siegler2005}.}
	\tablenotetext{g}{Entries for effective temperature and bolometric luminosity use polynomial relations from \citet{Filippazzo2015}.}
	\tablenotetext{h}{Although no period is yet observed in the radio, a period from photometric variability yields 1.86 $\pm$ 0.02 hr \citep{Harding2013a}.}
	\tablenotetext{i}{Although no period is yet observed in the radio, a period from photometric variability yields 8.9 hr \citep{Gizis2016}.}
	\tablenotetext{j}{Objects have been observed to display multiple highly circularly polarized pulses, but a periodicity has not yet been determined.}
	\tablenotetext{k}{Although no period is yet observed in the radio, a period from photometric variability yields $\sim$ 2.4 hr \citep{Artigau2009}.}
	\tablenotetext{l}{Follow-up observations since its initial radio detection have not revealed emission at the tentative period from the discovery paper \citet{Williams2017}.}
	\tablenotetext{}{R\textsc{eferences.} -- (1) \citet{Artigau2006}, (2) \citet{Berger2002}, (3) \citet{Berger2006}, (4) \citet{Bessell1991}, (5) \citet{Burgasser2005a}, (6) \citet{Burgasser2005b}, (7) \citet{Burgasser2006}, (8) \citet{Burgasser2013}, (9) \citet{Burgasser2015b}, (10) \citet{Burgasser2015a}, (11) \citet{Burgasser2015c}, (12) \citet{Castro2013}, (13) \citet{Clarke2008}, (14) \citet{Cruz2003}, (15) \citet{Cruz2007}, (16) \citet{Dupuy2010}, (17) \citet{Enoch2003}, (18) \citet{Filippazzo2015}, (19) \citet{Forveille2005}; (20) \citet{Freed2003}, (21) \citet{Gizis2000}, (22) \citet{Gizis2002}, (23) \citet{Gizis2011}, (24) \citet{Gizis2013}, (25) \citet{Gizis2015}, (26) \citet{Gizis2016}, (27) \citet{Harding2013a}, (28) \citet{Kao2016}, (29) \citet{Khandrika2013}, (30) \citet{Kirkpatrick2011}, (31) \citet{Koen2013}, (32) \citet{Lee2010}, (33) \citet{Legget2001}, (34) \citet{Lynch2016}, (35) \citet{McLean2012}, (36) \citet{Pineda2016a}, (37) \citet{Radigan2012}, (38) \citet{Reid2008}, (39) \citet{Reiners2010}, (40) \citet{Route2016a}, (41) \citet{Schmidt2007}, (42) \citet{Schmidt2015}, (43) \citet{Williams2014}, (44) \citet{Wilson2014}. }
\end{deluxetable*}

\section{Brown Dwarf Aurorae}\label{sec:aurorae}

Brown dwarf aurorae provide a natural explanation for many of the observational trends in UCD magnetism. The characteristics of the periodic radio pulses, that they are highly circularly polarized with large brightness temperatures, indicate that the ECMI must be the emission mechanism \citep[see][]{Treumann2006}, and that the aurorae, the consequence of energy dissipation from stable large-scale magnetospheric current systems, are present throughout the UCD regime. However, the physical conditions that power the aurorae remain an open question. Here, we discuss the implications of potential auroral scenarios on multi-wavelength brown dwarf emissions, how they fit within the observations of brown dwarf activity, and predictions based on the analogy with auroral systems in the gas giant planets of the Solar System. As a reference, we also compiled the physical properties and relevant observations of all of the potentially auroral UCD systems, based on their radio emission. This summary is presented in Tables~\ref{tab:ucds} and~\ref{tab:ucds2}. 

We discussed the dominant auroral electrodynamic engines in Section~\ref{sec:auroproc}. In the case of brown dwarfs, the three engines, stellar wind, satellite induced and co-rotation breakdown, take on slightly different forms. Although we expect brown dwarfs in binary systems orbiting much larger stars to interact with the companion's stellar wind in much the same way planets do around the Sun, for these brown dwarfs in large orbital separations ($\sim$10s of AU), the production of significant aurorae is disfavored \citep{Zarka2007}, while objects that may be sufficiently close are difficult to observe due to their small angular separation on the sky, requiring adaptive optics and/or very long baseline interferometry. Nevertheless, observationally many of the significant auroral brown dwarfs are isolated systems in the field. Work by \citet{Schrijver2009} and \citet{Nichols2012} suggests that motion through the ISM and reconnection in the large magnetosphere would not be able to produce enough energy to power the auroral radio emission. Both of these sets of authors suggest that co-rotation breakdown can provide sufficient energy to power the aurorae, with \citet{Hallinan2015} further suggesting that a satellite interaction could also provide sufficient energy. We discuss these two scenarios in Sections~\ref{sec:bdcorot} and~\ref{sec:planets}, respectively. 

Regardless, an important ingredient in the emergence of auroral phenomena in brown dwarf systems is the presence of large-scale magnetic field topologies in some objects (see Section~\ref{sec:dynamo}). The extended dipolar fields provide the conditions that generate an auroral electrodynamic engine, with the stronger fields at large distances enabling the strong coupling between the ionosphere and the middle magnetosphere. In contrast to the suggestions put forward in \citet{Williams2014}, given the connection between aurorae and large-scale fields, the radio load (ECM) and X-ray quiet objects likely host large-scale fields, while radio quiet and X-ray flaring objects likely display strong small-scale fields (see Section~\ref{sec:star_rad}). However, although a necessary component, the magnetic field topology may not be the distinguishing feature of the auroral processes (see Section~\ref{sec:planets}).

\subsection{Co-rotation Breakdown}\label{sec:bdcorot}

For co-rotation breakdown, the magnetospheric current and energy dissipated is dependent on the differential velocity between the co-rotating magnetosphere and an extended plasma disk, the electron density of that disk and on the strength of the brown dwarf's magnetic field \citep{Cowley2001, Nichols2012}. Furthermore, these properties also set the location of the auroral oval with respect to the magnetic axis, and hence the particular flux tubes connected to the shearing layer \citep{Cowley2001}. Thus, the magnetic field strength and rotational velocity play a crucial role in the generation mechanism and energy dissipation in this scenario. This electrodynamic engine would predict a strong dependence on rotational velocities and magnetic field strengths, both of which are much larger in brown dwarfs as compared to Jupiter, and could lead to much stronger aurorae in brown dwarfs \citep{Hallinan2015}. As seen in Figure~\ref{fig:rad_frac}, the radio detection fraction of UCDs rises in the more rapidly rotating objects, consistent with rapid rotation, P$\lesssim$3-4 hr (see Figure~\ref{fig:rad_frac}),  as a potentially critical ingredient to the generation of strong radio emission. The strength of stable auroral current systems also depends on the conductivity of the brown dwarf ionosphere \citep{Cowley2001, Nichols2012}. Although there are many variables involved, including the self-ionization from the electron-beam, this would indicate a dependence on T$_{\mathit{eff}}$ for the strength of auroral emissions. There is some hint of this in the quiescent radio emission of the auroral targets in Table~\ref{tab:ucds}, but larger samples are needed. Additionally, co-rotation breakdown predicts that auroral atmospheric emission should be predominately confined to a narrow oval around the magnetic axis; we discuss these atmospheric effects and their observability in Section~\ref{sec:atmo}.

However, a significant unknown for brown dwarfs is the existence of an extended equatorial plasma sheet that seeds the magnetospheric currents with plasma. Many possibilities for loading the magnetosphere with plasma have been put forward, including interaction with the ISM, reconnection events at the photosphere early in the brown dwarf's evolution and ablation of the atmosphere due to auroral currents \citep{Hallinan2015}. As a comparison, in the Jovian system, Io's volcanic activity provides the material to load the plasma disk. Analogous systems have not yet been confirmed around very low-mass stars and brown dwarfs; however there have been recent detections of small potentially rocky planets around UCDs with significant tidal interactions \citep{Udalski2015,Gillon2016,Gillon2017}. Moreover, models of planet formation around very-low mass stars suggests that these planets can form frequently and are likely to have radii similar to that of Earth \citep{He2017, Alibert2017}.

\subsection{The Possible Role of Planetary Companions}\label{sec:planets}

Given these new planetary detections, the satellite induced scenario for the electrodynamic engine becomes an intriguing possibility. The satellite flux tube interaction is generated by the differential motion of an orbiting satellite through the brown dwarf's magnetosphere. In this scenario the auroral power is a function of this velocity, the magnetic field strength at the orbital location of the satellite, and the cross sectional area of interaction of the planet, usually defined by the size of the obstacle \citep{Zarka2007}. For a rocky planet without an intrinsic magnetic field, the cross section is determined by the planet's exo-ionosphere (e.g., Io), whereas for a system with it's own magnetosphere the area is determined by the magnetopause distance of the planet within the brown dwarf's magnetosphere \citep[e.g., Ganymede,][]{Zarka2007}. Thus, stronger emission is expected for closer-in planetary systems, faster rotators, stronger magnetic field brown dwarfs, and larger more magnetized planets. Although many of these physical properties are unknown, using typical values, \citet{Hallinan2015} note that a planet orbiting in a $\sim$1.25 d orbit could generate enough energy to power auroral UCD emissions. Interestingly, under this scenario, the auroral emission may provide a separate constraint on the satellite radius, and provide a way to potentially confirm the existence and strength of the planetary magnetic field. Regardless, measurements of the surface feature and the emissions (see Section~\ref{sec:atmo}), associated with the flux tube can provide a way to distinguish the satellite scenario from co-rotation breakdown.

Nevertheless, because the origin of the magnetospheric plasma is an open question, the presence of planetary companions to brown dwarfs could be an important ingredient in both the satellite and co-rotation breakdown scenarios, which may additionally overlap in any given system. If the magnetospheric plasma is generated by some other means, then co-rotation breakdown may be the dominant brown dwarf auroral engine. While the discovery of a planet directly tied to auroral emissions in a brown dwarf would provide strong evidence for the importance of the satellites, an examination of the statistics of auroral objects and the occurence rate of close-in planets around them may provide important clues regarding the underlying auroral mechanisms. If the occurence rate of planets around auroral UCDs is significantly larger than the rate for the whole population of rapidly rotating UCDs, these data would be suggestive of the important role planetary companions play in auroral UCD systems. Currently, \citet{He2017} has showed that the occurrence rate of rocky planets around brown dwarfs with orbital periods less than 1.28 d is $\lesssim$67\%, and based on the recent planetary transit detections may be $\sim$27\%. This is at least consistent with a lower limit of $\sim$10\% from auroral radio and H$\alpha$ emission, if all of these brown dwarfs hosted unconfirmed planetary systems. When making this comparison, however, it may be important to also take into account the planetary system architectures. While a single very close-in planet may suffice to either seed the plasma disk or provide the obstacle for the auroral flux tube, tidal interactions may spin down the brown dwarf host, possibly limiting its ability to power significant aurorae, although this would depend on substantial unknowns, such as the degree of brown dwarf tidal dissipation \citep{Ribas2016, Gillon2017}. In contrast, interactions in multiple-satellite system, such as the one around Jupiter, can prevent the total tidal synchronization of the close-in satellites through resonant orbital interactions and a pumping of the orbital eccentricity \citep[see][and references therein]{Peale1999}. These effects could sustain significant tidal interactions throughout the system age that contribute to the internal satellite heat-flux and the generation of prominent volcanism, and consequently a plasma disk, without significantly spinning down the host brown dwarf; for example, the Jovian rotational period is $\sim$10 hr \citep{Bagenal2014}. If planetary companions are a necessary condition, these additional consideration may drive the difference between the overall UCD planet occurence rate and the prevalence of UCD auroral emissions. Current and future surveys, such as TRAPPIST and \textit{TESS}, will provide a better understanding of planet statistics in the UCD regime with which to compare the aurorally active set of brown dwarfs \citep{Gillon2011,Ricker2014}. 

An important example potentially illustrating the need for these distinct conditions to drive the electrodynamic engine, whether induced by satellite interactions and/or co-rotation breakdown, is the binary NLTT~33370. This object, composed of two nearly identical M7 dwarfs (see Table~\ref{tab:ucds}) at a separation of 2.528 AU (146.6 mas on the sky; \citealt{Dupuy2016}), has been studied extensively across different wavelengths to characterize its magnetic emissions \citep{Williams2015c}. This object shows the highly circularly polarized radio pulses characteristic of the ECMI; however, radio emission is observed in only one of the two components, the secondary, as shown in VLBI imaging \citep{Forbrich2016}. The two binary components also appear to have similar rotational velocities ($\varv \sin i \sim 45$ km s$^{-1}$), consistent with the radio period and the photometric period measurements \citep{McLean2011,Williams2015c, Forbrich2016}. That these two nearly identical objects, with the same mass, age, luminosity, effective temperature, and rotation period, display such different radio emission properties is difficult to explain without a significant underlying difference in the fundamental conditions which generate the radio emission. These observations can be explained by an auroral electrodynamic engine operating around NLTT~33370 B and not NLTT~33370 A. \citet{Forbrich2016} further use their astrometric monitoring to rule out Jupiter mass companions in the NLTT 33370 AB system in a wide variety of orbits; however a smaller rocky planet may be sufficient to provide the conditions to drive an auroral engine. Similarly the L0.5+L1.5 binary 2MASSI~J0746425+200032 with separation 2.9 AU (237.3 mas on the sky; \citealt{Konopacky2010}), shows periodic radio and H$\alpha$ emission at the same 2.072 hr period, attributable to the secondary, whilst the primary has a confirmed photometric period of 3.32 hr \citep{Berger2009, Harding2013b}. Like NLTT~33370, the components of 2MASS~J0746+2000 have similar physical properties but very different magnetic emissions. Planet searches around these objects would be potentially very interesting. An earth-mass planet in a circular 1 d orbit around these objects for example would have a radial velocity semi-amplitude up to $\sim$3.5 m s$^{-1}$. RV signals like these will be looked for in late M-dwarf targets by upcoming NIR high-resolution spectrographs, such as the Habitable-Zone Planet Finder \citep{Mahadevan2012}. Further study of these and other benchmark objects will be key in disentangling the nature of the auroral processes.

\subsection{Impact on Atmosphere}\label{sec:atmo}

If the ECMI is present, then a strong field-aligned current is driving an electron beam to precipitate into the atmosphere. This has a significant impact on the brown dwarf atmosphere, leading to the creation of multi-wavelength auroral emissions. The same physical processes that take place in the Jovian system should operate in brown dwarfs (see Section~\ref{sec:auroproc}), however important differences may arise due to the different atmospheric conditions and potentially due to the different properties of the auroral electron beams. 

\subsubsection{Emission Line Features}\label{sec:atmo_em}

The collision of high energy electrons with an atmosphere predominantly of neutral H/H$_{2}$, as is found in brown dwarfs, leads to the ionization and excitation of the hydrogen gas. The de-excitation and recombination of the hydrogen will lead to considerable Balmer and Lyman series emission, as well as UV emission in the Lyman and Werner bands \citep{Badman2015}. Consequently, brown dwarfs hosting auroral radio emission are likely to generate surface features which are bright in H$\alpha$, Ly$\alpha$ and the FUV. Thus, the presence of these emissions should be correlated with the presence of ECM radio emission. The statistics of radio detections and H$\alpha$ emission in late L dwarfs and T dwarfs show evidence for this correlation \citep{Kao2016, Pineda2016a}. The auroral FUV emission in brown dwarfs has not yet been discovered, but would be an important confirmation of the effects of auroral electron beams in brown dwarf atmospheres. Moreover, following the Jovian example, this FUV emission would be diagnostic of the total auroral energy and is sensitive to the auroral electron energy distribution \citep[][Pineda et al. in prep]{Bhardwaj2000,Badman2015}.

Additionally, the nature of the excitation mechanism for the hydrogen emission features could potentially be used as a diagnostic of the auroral process. Detailed modeling is required to predict the Balmer line spectrum from the electron impacts on the brown dwarf, however the emission ratios of the lines likely deviate from expectations from case B recombination and simple LTE gas models. An analysis of the Balmer series emission line ratios (decrement) by \citet{Stelzer2012}, for one of the late M dwarfs with strong radio emission showed that these models proved to be a poor fit to the data and did not follow the expectations from a standard stellar chromospheric perspective. An auroral NLTE origin for the Balmer series emission could account for this discrepancy. More detailed observations of the emission lines of auroral brown dwarfs would aid in this endeavor.

\begin{figure*}[tbp]
	\centering
	\includegraphics[width=0.9\textwidth]{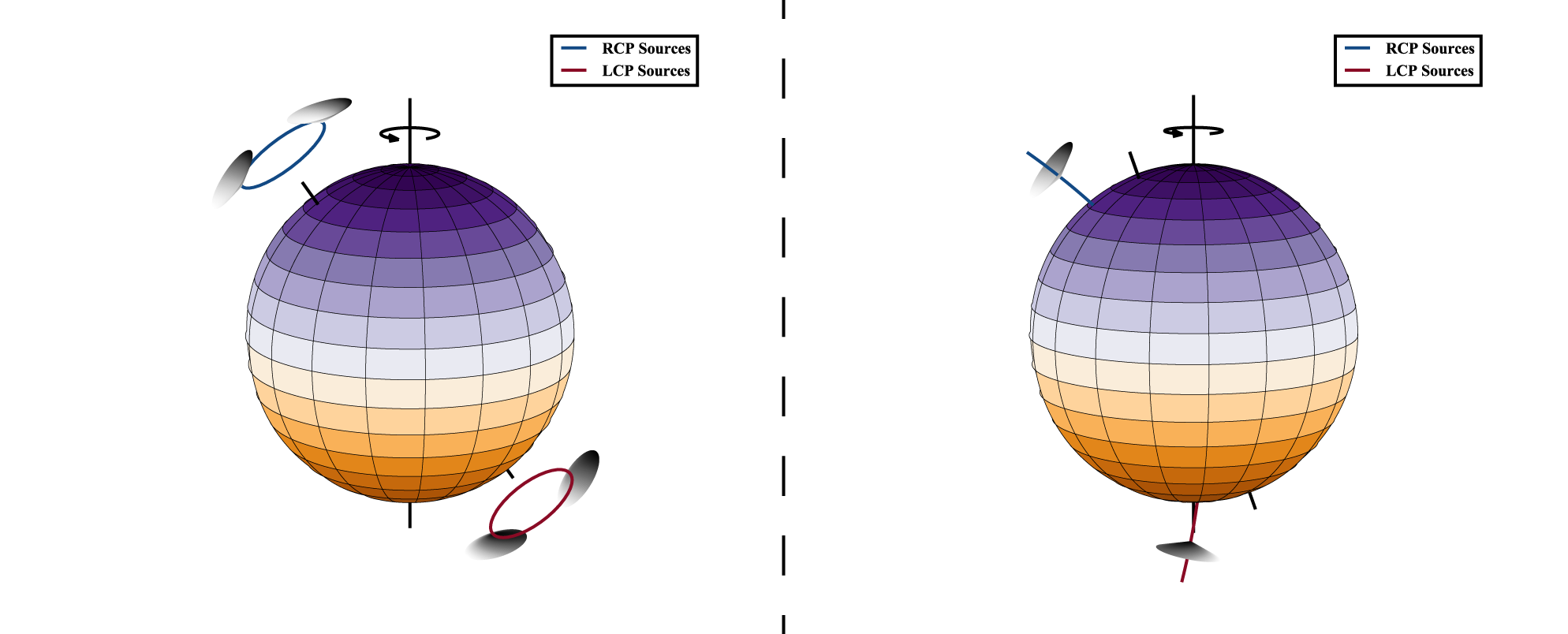} 
	\caption{Visualizations of the auroral radio sources, see Section~\ref{sec:beam}, with example emission cones. \textit{Left} - Geometric configuration of an auroral ring with right circularly polarized sources in the northern hemisphere and left circularly polarized sources in the southern hemisphere, viewed at an inclination of 75$^\circ$ with a magnetic axis tilted 35$^{\circ}$ from the rotational axis. The beaming patterns of Figures~\ref{fig:beampattern_20} and \ref{fig:beampattern_35} are generated by sources like these. \textit{Right} - Geometric configuration for auroral sources along a particular flux tube, as in the satellite-induced scenario, viewed at an inclination of 80$^{\circ}$ with a magnetic axis tilted 20$^{\circ}$ from the rotational axis. The beaming patterns of Figure~\ref{fig:satellitepattern} are produced by these auroral sources.}
	\label{fig:sources3D}
\end{figure*}

Depending on the auroral electrodynamic engine, these features will appear as ovals around the magnetic axis or be localized to the satellite flux tube foot point, which have different observational signatures. The different morphologies of the surface features will be imprinted on the emission lines, like H$\alpha$. In contrast to chromospheric emission which will be fully rotationally broadened, the line profile of the auroral H$\alpha$ emission will be determined by the size and location of the auroral surface feature. Emission localized at a flux tube foot point would be narrow and centered at velocities set by the longitude of the surface feature. The emission features of an auroral oval, would be wider and span a range of velocities consistent with its L-shell \footnotemark[3] and the tilt of the magnetic axis relative to the rotational axis. The shape of the line would also deviate from a simple velocity broadened Gaussian profile because the emission at the center of the oval would be missing. High-resolution spectroscopic observations of auroral brown dwarf emission features can be used to reconstruct these properties of the surface features (Pineda et al. in prep).

\footnotetext[3]{The L-shell is the distance, in units of the object's radius, along the magnetic equator which maps to a location on the stellar surface along the corresponding field line traversing the magnetic equator.}

The observed variability also depends on the auroral engine. For an auroral oval, the feature rotates around as the magnetic axis rotates around. Depending on the viewing geometry, the auroral surface emission could rotate in and out of view creating a sinusoidal signal, like what is observed on LSR J1835+3259 \citep{Hallinan2015}. This variability is also imprinted on the velocity profile of the emission lines. In the case of a flux tube foot point the time variability of the emission is additionally modulated by the orbital motion of the satellite in addition to the rotational period of the object. This leads to long term variability of the emission features. Emission line variability in excess of the rotational period has already been confirmed in one auroral brown dwarf \citep{Pineda2016a}. Like the Jovian system, the auroral emission features are likely to be intrinsically variable on short timescales, reflecting changes in the electron beam energy distribution and auroral current system. Stellar chromospheric emission is also characterized by intermittent variability and can show enhanced emission within star spot regions \citep{Reiners2008,Lee2010}.

As discussed in Section~\ref{sec:auroproc} the ionization of molecular hydrogen from the electron impacts leads to the creation of ionized triatomic hydrogen. H$_{3}^{+}$, however, is sensitive to the conditions in the atmosphere, and could be an important diagnostic with strong emission features at 2 $\mu$m and 4 $\mu$m \citep[][]{Tao2011,Tao2012,Badman2015}. The ro-vibrational features are thermally excited and depend on the atmospheric temperature. Additionally, the presence of H$_{3}^{+}$ is limited by the electron number density and the concentration of gas species, like CH$_{4}$ and H$_{2}$O that act to destroy the ion \citep[][]{Badman2015}. Interestingly, although auroral electron beams in brown dwarfs are likely to create H$_{3}^{+}$, the concentration may not build up significantly if the ion is quickly destroyed. This would occur for large ionization fractions (higher in brown dwarfs relative to Jupiter but still mostly neutral, see Section~\ref{sec:chromo}), or a beam with high mean electron energies that penetrates to the deep atmospheric layers, below the homopause, where the concentration of molecules is high. Observations of these features could potentially be used to constrain the brown dwarf auroral beam electron energy distribution (Pineda et al. in prep). If the auroral brown dwarfs do host significant H$_{3}^{+}$ emissions, it may be possible to observe them with the upcoming \textit{James Webb Space Telescope}. This could also provide an additional diagnostic that would distinguish between an aurorally active UCD and one hosting a chromosphere/corona.

\begin{figure*}[tp]
	\vspace{-0.8in}s
	\centering
	\includegraphics[width=0.9\textwidth]{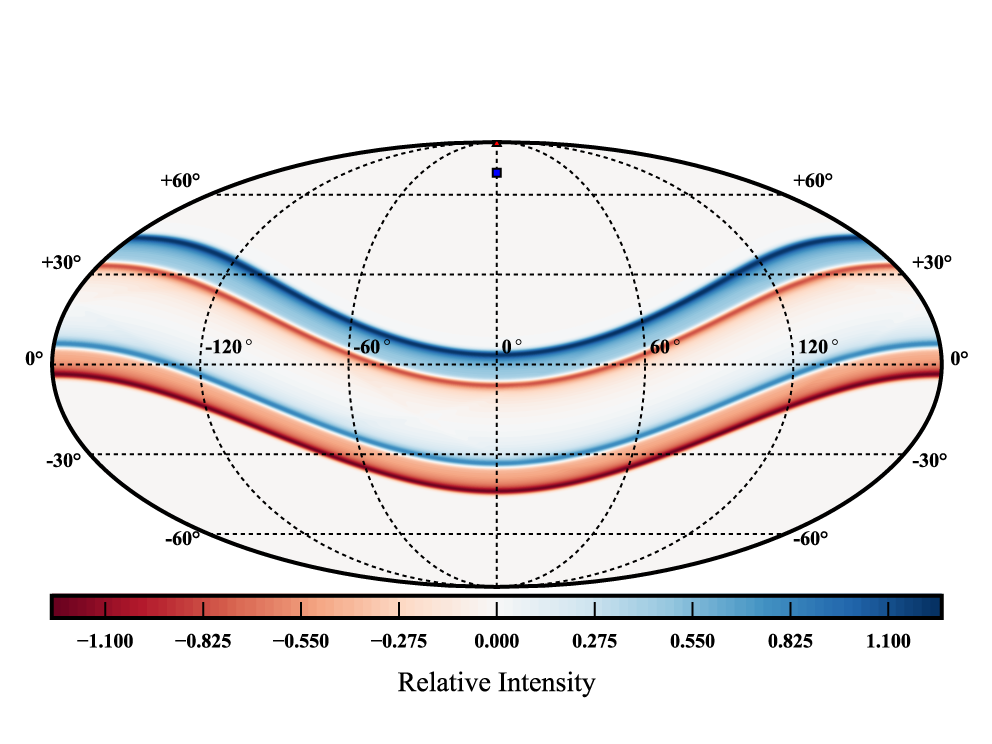} 
	\caption{The beaming pattern of auroral radio emission in Stokes V for a single frequency channel in a uniform auroral ring, as seen from the emitting object (see~\ref{sec:beam}). The emission is normalized with positive values of the intensity corresponding to right circular polarization and negative to left circular polarization. The emission model assumes that ECMI radio sources are located in a continuous ring around the magnetic axis near the stellar surface and that each emits in a hollow cone with an opening angle of 85$^{\circ}$ with a cone width of 2$^{\circ}$. The sources are placed at an L-Shell of 30, and $\sim$12.5$^{\circ}$ from the magnetic axis. The magnetic axis is further defined to be 20$^{\circ}$ from the rotation axis, with its direction indicated by the blue square on the plot. As the object rotates, this beam pattern rotates to the right on a fixed sky.}
	\label{fig:beampattern_20}
\end{figure*}

\begin{figure}[htbp]
	\centering
	\includegraphics{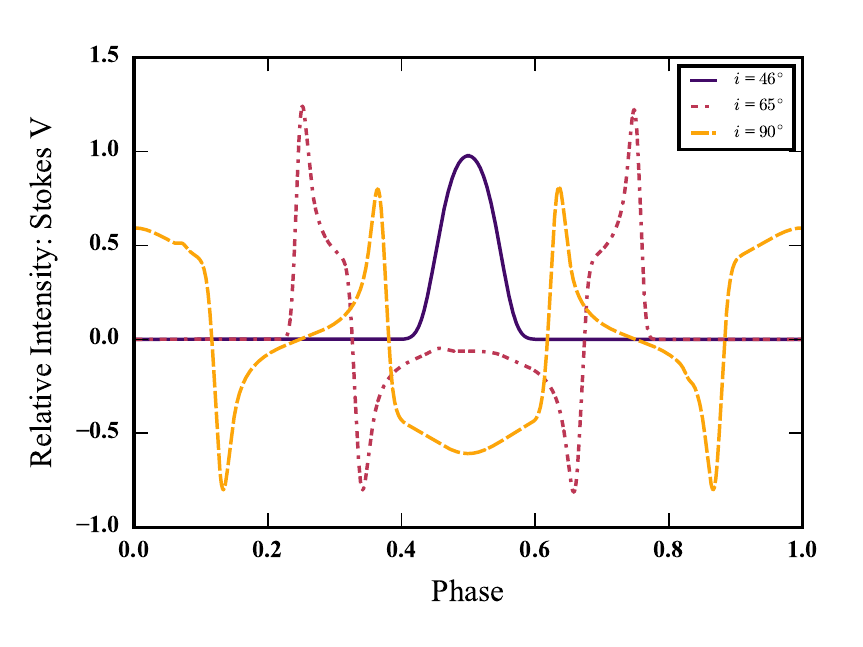} 
	\caption{Intensity light curves in Stokes V from the variable ECM radio emission beam pattern of Figure~\ref{fig:beampattern_20}, as viewed with different object inclinations. These light curves correspond to latitudinal cuts in the normalized intensity pattern created from the rotational variation of the emission. Phase 0 corresponds to longitude 0$^{\circ}$ in Figure~\ref{fig:beampattern_20}. The beam pattern can generate a broad variety of light curves depending on the relative geometry of the object and the observer. }
	\label{fig:bp_lightc}
\end{figure}

Because these emission features depend on the atmospheric conditions, like the Jovian system, their intensity is diagnostic of the upper atmospheres of brown dwarfs. If detected, features like the H$_{2}$ Werner band emission and H$_{3}^{+}$ lines would provide the only probe of these atmospheric regions well above the photosphere, however the effect will require detailed modeling to understand thoroughly.

\subsubsection{Photometric Variability}\label{sec:bd_var}

The impact of the electron beam on the atmosphere deposits large amounts of energy, heating the atmosphere. In Jupiter, the bulk of the auroral energy emerges as thermal emission (see Section~\ref{sec:auroproc}). The effect of this heating will depend on the atmospheric layers, where the bulk of the energy deposition takes place. Studies examining the effect of thermal perturbations in brown dwarf atmospheres have illustrated the wavelength dependence of the variable emission and demonstrated that stronger variability at IR wavelengths is generated when the energy is deposited higher in the atmosphere \citep{Robinson2014,Morley2014}. In addition to changing the temperature profile, the auroral energy alters the chemical structure and can impact the opacity of the auroral surface region. \citet{Hallinan2015} argues that this may be the mechanism generating the auroral surface features seen in broadband optical monitoring of auroral brown dwarfs \citep{Harding2013a}.

\begin{figure*}[tbp]
	\vspace{-0.8in}
	\centering
	\includegraphics[width=0.9\textwidth]{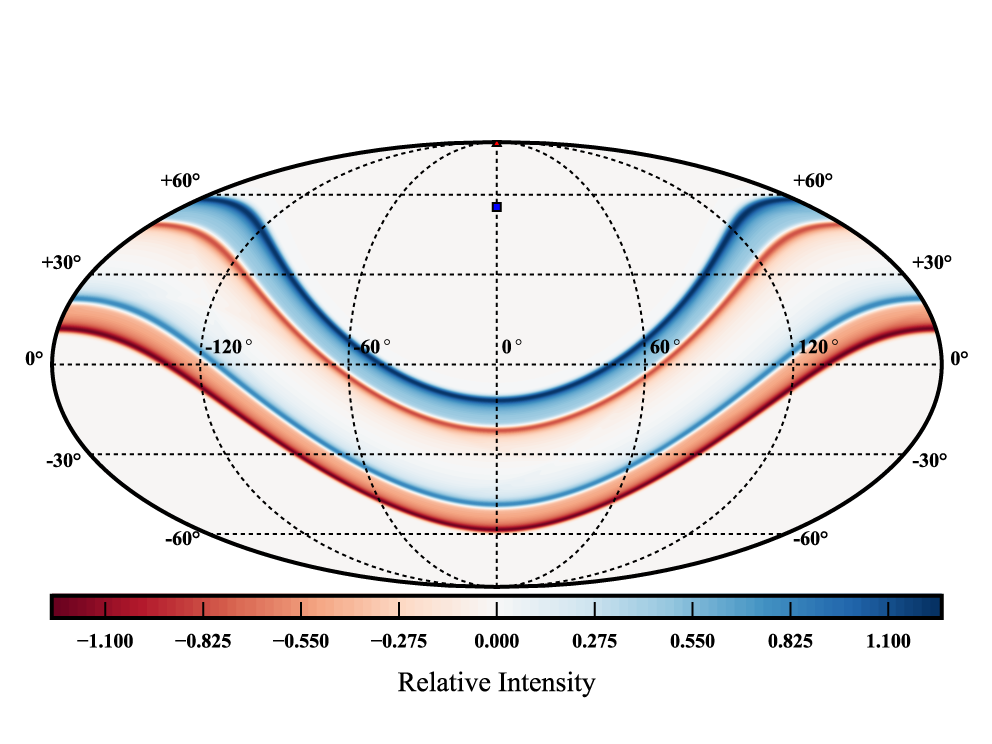} 
	\caption{Same as Figure~\ref{fig:beampattern_20}, but with the tilt of the magnetic axis set to 35$^{\circ}$. With a larger angle between the rotational and magnetic axes, the proportion of the sky which is traversed by the auroral beam pattern increases.}
	\label{fig:beampattern_35}
\end{figure*}

Modeling the auroral electron beam impact is key to understanding these effects, however there are several expectations with regards to the photometric variability of auroral surface regions. The variability is generated near the magnetic axis and may be preferentially located near the rotational axis, generating sinusoidal variations as the feature moves into and out of view. The surface features are relatively steady and remain present on long time scales, exceeding the transient signals expected individually from star spots and/or variable cloud structures. These features can potentially be distinguished from long-lived storm systems (akin to Jupiter's great spot) through their latitudinal distributions; the storms are more prevalent along the equatorial belts, whereas the auroral variability is likely at high latitudes near the rotational axis. There should also be a strong correlation between objects displaying auroral radio emission and those showing long-term periodic variability. Comparing those objects with large NIR variability, J and K bands, to those with auroral radio emission and/or H$\alpha$ emission shows a mixed record. The canonical L/T transition variables 2MASS~J21392676+0220226 and SIMP~J013656.5+093347.3, for example, do not display H$\alpha$ emission. This is somewhat surprising, since SIMP~0136+09 was detected with highly circularly polarized radio emission; however this may point toward the influence of a satellite (see Section~\ref{sec:atmo_em}; \citealt{Kao2016, Pineda2016a}). Other objects like Luhman16 do show photometric variability and no sign of auroral activity \citep{Osten2015}. Some objects with IR variability from \textit{Sptizer}, on the other hand, do show activity, like 2MASS~J00361617+1821104 \citep{Metchev2015}. Long-term monitoring of these objects may be needed to distinguishing the mechanisms producing the surface features, but there does not appear to be a one-to-one connection between the auroral activity and photometric variability \citep[see also][]{MilesPaez2017}. The data are consistent with two independent effects taking place on similar populations of objects, with general variability more common than auroral activity, and their interplay being responsible for the observed phenomena in some objects throughout the UCD regime \citep{Croll2016a}.

\subsection{Auroral Beaming Geometry and Radio Emission}\label{sec:beam}

The different proposed electrodynamic engines of auroral emission produce distinct geometric beaming patterns. To understand this better we modeled the radio emission patterns for both a uniform auroral ring and a single auroral flux tube, to approximate the ECMI radio sources generated by co-rotation breakdown and the magnetosphere-satellite interaction, respectively. We visualize these radio sources in Figure~\ref{fig:sources3D}. The left panel of the plot shows the locations of the auroral radio rings, situated around the magnetic axis in both the northern and southern hemispheres, while the right panel shows the emitting field lines for a satellite flux tube. Overlaid are illustrative hollow radio emission cones (see Section~\ref{sec:auroproc}). At each point along the blue and red lines of the plots in Figure~\ref{fig:sources3D}, we consider there to be an ECMI radio emission source, and model the resulting cumulative beam pattern on the sky \emph{as seen from the brown dwarf}. 

Modeling the brown dwarf radio emission on what is observed from Jupiter, we can use some basic properties of the emission to determine what to expect in the beaming pattern and the variability from brown dwarfs. The Jovian radio aurorae are beamed into thin ($\sim$1-2$^{\circ}$) hollow cones, emitting in directions nearly perpendicular to the local magnetic field direction, with cone half-angles $\sim$80-90$^{\circ}$ \citep{Treumann2006}. The emission is highly circularly polarized and appears to show different directions of polarization from the different magnetic poles, consistent with X-mode waves in the plasma \citep{Zarka1998}. We take the large-scale field to be predominantly dipolar and use it to map the acceleration regions to different locations in the magnetosphere. With these assumptions, we can examine the expected beam patterns for different systems and radiation sources, and what effect these generic properties have on the observations of auroral emission. By contrast numerous authors have used these similar assumptions to try to directly constrain the properties of individual objects through models of their dynamic spectra \citep{Yu2011,Lynch2015,Leto2016}. Although interesting, these efforts require many assumptions of typically uncertain parameters for a problem that is difficult to answer accurately, even with much more information, as has been attempted with radio observations of Jupiter (see \citealt{Hess2011}).

\begin{figure}[tbp]
	\centering
	\includegraphics[width=0.5\textwidth]{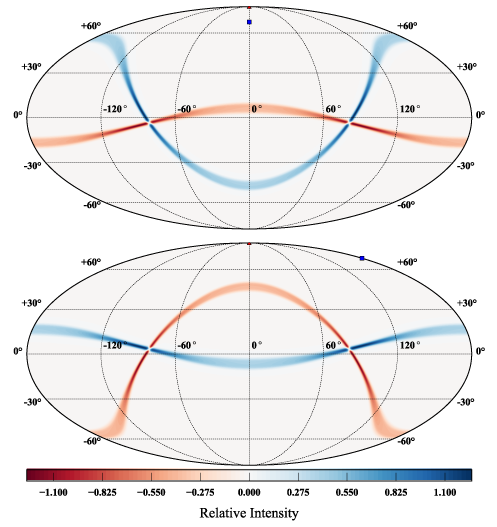} 
	\caption{Emission beam patterns on the sky, as in Figure~\ref{fig:beampattern_20}, for a set of ECMI radio sources at a fixed longitude associated with a current system fixed to a brown dwarf satellite flux tube. The range of sources span emission frequencies 3-12 GHz, for a dipolar field strength of 4.5 kG at the brown dwarf surface along the magnetic axis. The L-shell is set to 10 and the magnetic axis is tilted by 20$^{\circ}$. The top plot is oriented with the magnetic axis, indicated by the blue box, pointing toward the direction of the planet and the bottom plot differs in phase by 180$^{\circ}$, with the axis pointing away from the planet. The emission pattern transforms continuously on the sky with the rotation and orbital motion of the satellite. }
	\label{fig:satellitepattern}
\end{figure}

\begin{figure}[tbp] 
	\centering
	\includegraphics[width=0.5\textwidth]{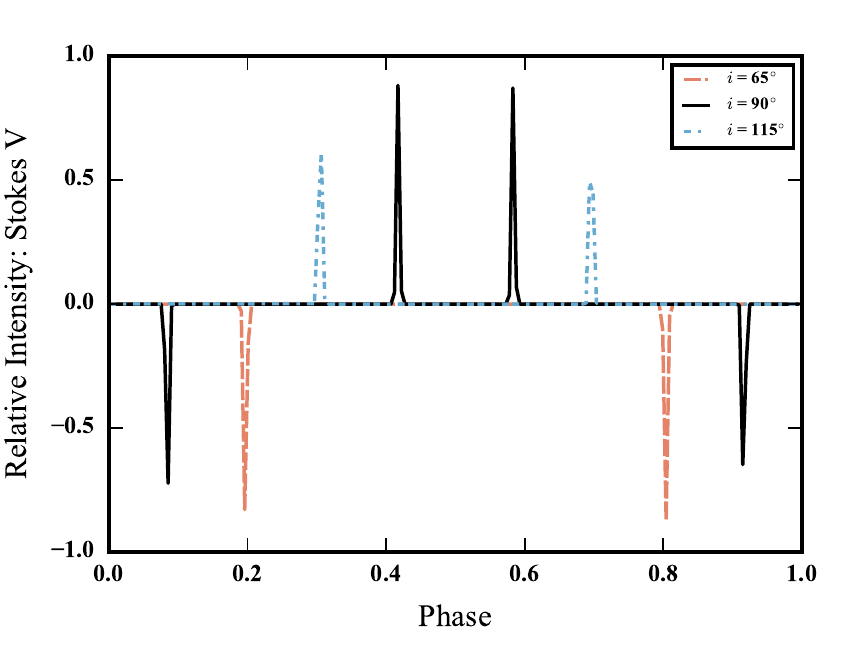} 
	\caption{Radio light curves of single frequency channels in Stokes V intensity with the object's rotation for different observers viewing a satellite induced auroral radio source, as in Figure~\ref{fig:satellitepattern}. The observers see narrow pulses when the beam pattern traverses their line of sight, and depending on the viewing geometry, the observer may see only one polarization or the other, or both polarizations coming from the north and south hemispheres respectively. $i >90^{\circ}$ corresponds to the object's rotational axis pointing away from the observer.}
	\label{fig:satrad_LC}
\end{figure}

In the case of co-rotation breakdown, the ECMI sources are expected to be concentrated in a ring, near the object's surface, around the magnetic axis, with the location defined by the L-shell associated with the magnetospheric current system. In Figure~\ref{fig:beampattern_20}, we show a model beam pattern on the sky, in longitude and latitude, as viewed from the brown dwarf in a two-dimensional projection. The emission sources are from an auroral ring in the north and south hemispheres, similar to the left panel of Figure~\ref{fig:sources3D}. The pattern shows the intensity as a function of position on the sky, using both right-hand circular polarization (positive intensities) and left-hand circular polarization (negative intensities). The inclination is set to 90$^{\circ}$ with the magnetic axis tilted at an angle of $20^{\circ}$ in a uniform ring close to the surface (1.4 R$_{*}$) with an L-shell of 30, corresponding to sources located $\sim$12.5$^{\circ}$ from the magnetic axis. The pattern of emission encompasses a large swath of the total sky. The peak emission intensities are generated at the edge of the beam pattern where the contributions from sources at multiple points along the ring contribute constructively. As the object rotates, this pattern traverses the sky in longitude and generates the periodic variability of ECM emission.

For the observer, they would see very different light curves depending on their geometrical orientation with respect to this beaming pattern. As plotted in Figure~\ref{fig:beampattern_20}, the pattern uses an inclination of 90$^{\circ}$, and the observer sees the light curve generated by the horizontal cut of this pattern at 0$^{\circ}$ latitude. However, other inclination light curves can be read directly from these figures by looking at the latitudinal cut corresponding to $90 - i$. We plot some example light curves in Figure~\ref{fig:bp_lightc}, for a selection of viewing inclinations from Figure~\ref{fig:beampattern_20}. If the object is viewed at an inclination of 0$^{\circ}$, viewed above the rotational axis, there is no ECM emission visible. Consequently, the beaming pattern and the viewing geometry can have a significant impact on the detection statistics for radio surveys of brown dwarfs. At other inclinations, the variability encompasses multiple pulses, a single pulse and/or multiple polarizations. Thus, a variety of light curve morphologies are possible and can change depending on the particular emission parameters, like cone width and cone opening angle. A narrower cone width, for example, would produce sharper pulse features. Comparing the morphology to the radio light curves produced in various surveys, we see a broad similarity between these shapes and the generic light curves of our model, based on basic assumptions about the emission process \citep[][]{Hallinan2008,Berger2009,Williams2015a,Kao2016}. The beam pattern in Figure~\ref{fig:beampattern_20}, uses uniform rings of equal intensities, but observations of Jupiter indicate that the auroral emission can fluctuate (see \citealt{Zarka1998}); moreover, the rings may not be populated uniformly with sources, leading to gaps in the light curves and variable pulse intensities and asymmetries between the right and left circularly polarized emissions. Thus, although an auroral ring may be predicted to produce multiple detectable pulses of both polarizations in a single period, they may not always be visible. However, if they are seen, then the separation of the pulses can be used to constrain the cone opening angle and the orientation of the magnetic axis relative to the rotational axis, keeping in mind the various degeneracies of doing that inversion.

An examination of these beaming patterns, reveals that they encompass a large fraction of the sky as the object rotates around. Moreover, the fraction is larger when the magnetic axis is misaligned with respect to the rotational axis. In Figure~\ref{fig:beampattern_35}, we show a beam pattern similar to that of Figure~\ref{fig:beampattern_20}, but with a magnetic axis 35$^{\circ}$ misaligned instead of 20$^{\circ}$, where the pattern encompasses 2/3 of the sky. The proportion depends on the L-shell of the radio sources, the source height and cone opening angle, but is large for reasonable values of these parameters, $\sim$50\%. \citet{Pineda2016a} and \citet{Kao2016} argue that the large overlap between objects with radio auroral emissions and optical auroral emissions indicates that the geometric selection effect biasing the radio detections may not be very strong. If this is the case, and the fraction of auroral objects is driven largely by the proportion of objects with physical conditions amenable to the generation of auroral magnetospheric currents, then generically, the magnetic axes of brown dwarfs are likely to be misaligned or at least a substantial component of the large-scale field is misaligned. Given these considerations, the detection of quiescent radio emission from a pole-on L dwarf is unlikely to generate pulsed radio emission unless the magnetic axis is totally misaligned \citep{Gizis2016}. Similarly, the detection of variable circularly polarized emission in the data of WISEP J112254.73+255021.5, has been used to argue for a misaligned magnetic axis in that T dwarf \citep{Williams2017}. This is particularly interesting because our model light curves predict, in these cases, some circularly polarized emission between the peak pulses with relatively uniform auroral rings.

In contrast to the auroral oval case, the generation of ECMI through currents in a flux tube connecting a satellite and a brown dwarf produces a very different beam pattern and variability signal. Under these circumstances, the radio source region is confined to the longitude associated with the flux tube of the satellite (see right panel of Figure~\ref{fig:sources3D}). However, the properties of the individual radio sources should be the same: wide and thin hollow cones with different polarizations in the northern and southern hemispheres. In Figure~\ref{fig:satellitepattern}, we show an example beam pattern for a satellite induced source region with L-shell of 10, spanning several heights analogous to multiple frequencies (e.g.~3 - 12 GHz), and two different instances of the rotational period, assuming the planet has remained relatively fixed at a longitude of 0$^{\circ}$. In the top panel, the direction of the magnetic dipole is pointing toward the satellite and in the bottom panel it is pointing away, a 0.5 difference in rotational phase. There are several effects illustrated in this diagram. The inclusion of several frequencies, in contrast to the single frequency beam patterns of Figures~\ref{fig:beampattern_20} and \ref{fig:beampattern_35}, broadens the pulses when examined in broadband light curves. Since the different frequencies map to different field strengths, the corresponding emission cones are pointed in slightly different directions, creating a slight broadening of the broadband ECMI pulse and slightly different arrival times for the emission at different frequencies as the emission cones sweep into view during the rotational period of the brown dwarf. We do not consider the effect of the frequency dependent refraction of the radiation, which generates more significant deflections at lower frequencies, as it propagates out of the ECM emission region \citep{Mutel2008}. Attempting to do so requires further assumptions about the plasma density in the magnetosphere, which has little observational constraints; however, including this physical effect would narrow the emission beam patterns (less instantaneous sky coverage) for the lower energy radiation, perhaps leading to observing frequency dependent detection statistics for radio auroral pulsations. More lower frequency observations ($\sim$1 GHz) are required to test this hypothesis. The large overlap in the auroral H$\alpha$ and radio detection statistics may also suggest that the refraction may not be very strong in the 4-8 GHz radio band.

Additionally, Figure~\ref{fig:satellitepattern} illustrates that when the magnetic axis is misaligned, the instantaneous beam pattern on the sky changes in shape and direction transforming continuously from the pattern in the top panel to the pattern in the bottom panel (for each fixed polarization) and back over the course of the brown dwarf's rotational period, traversing large portions of the sky, achieving a total sky coverage comparable to the auroral oval scenario. In effect, the pattern can be considered as a result of the flux-tube tied to the satellite traversing multiple L-shells and longitudes, doing a circuit with respect to the magnetic axis with the rotational period of the brown dwarf. These rotational effects do not occur, if the magnetic field is well aligned with the rotational axis. Thus, for the auroral satellite scenario the radio emission can produce periodic light curves at the brown dwarf rotation period with both single and multiple polarization peaks, depending on the relative geometry of the observer and the source. We show some example light curves demonstrating this effect for single frequency channels in Figure~\ref{fig:satrad_LC}. The different lines correspond to observers viewing the target at different inclinations, along the longitude 0$^{\circ}$ line of Figure~\ref{fig:satellitepattern}. These observers see narrow pulses as the beam pattern sweeps over their lines of sight during the rotational modulation of the brown dwarf. Additionally, the changes in the magnetic field strength at the location of the satellite also contribute to variability in the observed intensity of the radio pulses. Super-imposed on this rotational variation, the satellite scenario further predicts a modulation of the radio emission on the orbital period of the planet around the brown dwarf. Radio emission models of TVLM513-46546, might show some indications of this behavior \citep{Leto2016b}. The long-term orbital variations may be responsible for the observed changes in pulse polarization observations, alternatively interpreted as a potential indication of stellar cycles in UCDs \citep{Route2016c}. For a single object hosting both of these electrodynamic engines, co-rotation breakdown and a satellite flux-tube, depending on the relative viewing geometry and underlying parameters, the observer might intercept both, none or only one of the radio beam patterns of the ECMI emission. Long-term radio monitoring of these targets will provide a means to potentially measure these effects \citep[see also][]{Wolszczan2014}.

\subsection{Quiescent Radio Emission}\label{sec:pulseQ}

An important aspect of the detection of radio emission from UCDs, has been the distinction between the pulsed emission and a quiescent component at GHz frequencies. While the characteristics of the pulsed emission have identified it as due to the ECMI, the cause of the quiescent component is poorly understood. Based on the beam patterns and light curves from Section~\ref{sec:beam}, there could be ECMI emission between pulse peaks, although it would likely be much weaker and variable than the idealized emissions considered here. However, the polarization of the quiescent emission does not reflect a potential ECMI origin \citep[however, see][]{Williams2017}. UCD quiescent radio emission typically shows low levels of circular polarization, however the constraints have not been particularly stringent \citep[e.g.,][]{Kao2016}. Based on radio detections at $\sim$100 GHz, \citet{Williams2015b} showed that the quiescent emission of TVLM~513-46546, one of the benchmark targets with periodic pulsations, was consistent with synchrotron and/or gyro-synchrotron emission. Whether the energetic electrons are highly relativistic or only mildly relativistic has been difficult to assess. Consequently, the polarization of the quiescent component could be an important distinguishing feature. A gyro-synchrotron source would show some circular polarization, whereas a synchrotron source would display significant linear polarization \citep{Dulk1985}. The low degree of circular polarization argues against the ECMI and may place constraints on gyro-synchrotron, but the future confirmation of significant linear polarization will be an important indicator of the synchrotron source for the quiescent radio component and the relativistic nature of the energetic electron population.

\begin{figure}[tp]
   \centering
   \includegraphics{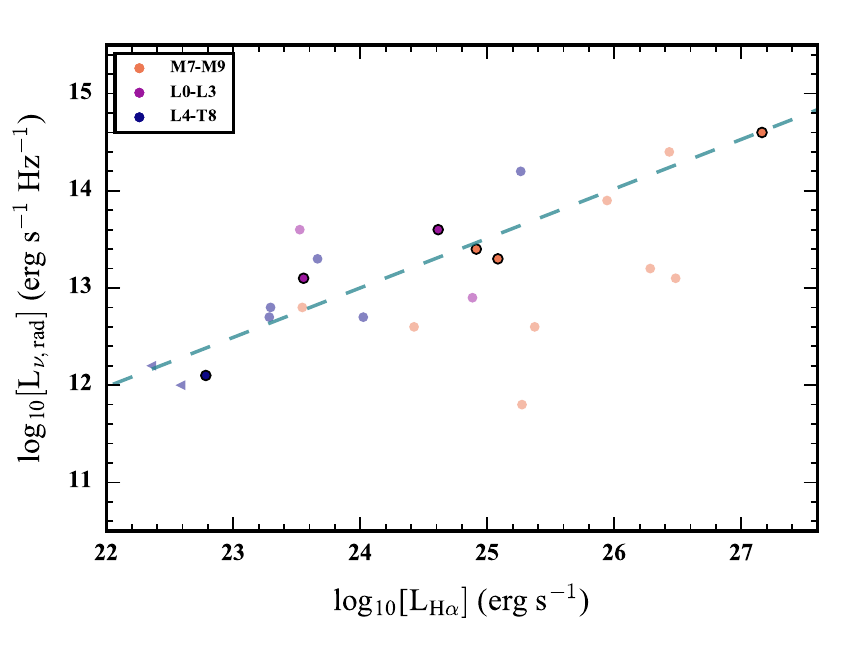} 
   \caption{The observed quiescent radio luminosity of radio UCDs plotted against their H$\alpha$ luminosities, taken from Tables~\ref{tab:ucds} and~\ref{tab:ucds2}. The objects with confirmed periodic pulsations (from Table~\ref{tab:ucds}), are outlined in black, showing a correlation between the two emission types (r = 0.95, p=0.004); a best fit regression line of slope 0.51 and ordinate intercept 0.78, for these objects, is plotted as a dashed line. This suggests there may be a connection between the auroral processes and the quiescent radio emission, see Section~\ref{sec:pulseQ}. The points are plotted in different shades according to their spectral types, as in Figure~\ref{fig:UCDVsin}, with limits indicated by triangles.}
   \label{fig:pulsedQ}
\end{figure}

This, however, raises the question as to how these brown dwarfs energize the electrons responsible for this relativistic emission. Stellar gyro-synchrotron emission is connected to the same heating and acceleration mechanisms that power the hot plasma of the coronae and stellar flares. However, the lack of X-ray emission (see Figure~\ref{fig:UCDXray}) suggests these processes are weak in the cool atmospheres of UCDs, especially in late L dwarfs and T dwarfs, many of which have quiescent radio detections \citep[e.g,][]{Kao2016}. Instead, one possibility for this emission, as first proposed in \citet{Hallinan2006}, is a brown dwarf counterpart to the synchrotron radiation belts of the Jovian system. Jupiter exhibits a large population of magnetospherically confined energetic electrons ($\sim$1 MeV) in its equatorial regions that generate strong synchrotron emission at decimeter wavelengths \citep[see][and references therein]{Bagenal2014}. If similar structures exist on these brown dwarfs, they could generate the quiescent emission at GHz and higher frequencies, given the comparatively stronger magnetic field strength of brown dwarfs relative to Jupiter. 

Whilst the cause of the quiescent emission remains an open question, there may be a clue as to its origins in its relation to the ECMI pulsed radio emission. There is a large overlap between the population of objects with periodic pulsed ECMI emission and those with a detected quiescent radio component \citep[e.g.][]{Hallinan2008,Kao2016}. Moreover, for many objects, initial radio detections come from short surveys of only 1-2 hr per object, which can be much less than the rotational periods, potentially missing the periodic nature of some of the bursts, or lacking sufficient sensitivity to probe quiescent emission \citep[e.g.][]{McLean2012, Route2013, Route2016b}. Follow-up efforts on these detections often confirm the presence of periodic pulsations on top of a quiescent radio background \citep[e.g.,][]{Route2012,Williams2015a}. Additionally, many of the sources listed in Table~\ref{tab:ucds2}, already show evidence for periodic pulsations, but await further follow-up observations for confirmation \citep{Kao2016}. This connection between the two kinds of radio emissions is consistent with the of presence brown dwarf synchrotron radiation belts. If this is indeed the source of the quiescent emission then the inner magnetospheric region must be loaded with energetic plasma; a requirement similarly critical for the generation of auroral currents. In the Jovian case, the major plasma source is the moon Io \citep[e.g.,][]{Bolton2015}. The material picked up from the moon is ionized and energized by the fast rotating magnetosphere, the ultimate source of energy for the system, leading to a confined population of highly energetic electrons in the inner magnetosphere \citep{Bolton2015}. The presence of this plasma in the large-scale magnetosphere may be the crucial link connecting the processes.

\footnotetext[4]{The data sample is small and thus the $p$ values associated with the Pearson correlation coefficient, $r$, are not particularly meaningful, but are included here for completeness.}

Interestingly, there also appears to be a correlation between the quiescent radio emission and H$\alpha$ emission strengths. In Figure~\ref{fig:pulsedQ}, we plot the quiescent radio luminosity against the H$\alpha$ luminosity for radio UCDs (taken from Tables~\ref{tab:ucds} and~\ref{tab:ucds2}). These data show a Pearson correlation coefficient of 0.54 (p=0.01)\footnotemark[4], indicating a positive correlation between the values. If we restrict the data to just the periodically pulsing sources from Tables~\ref{tab:ucds}, the Pearson correlation coefficient rises to 0.95 (p=0.004)\footnotemark[4]. Although there are not very many data points, they further indicate a possible connection between mechanisms producing the aurorae and the quiescent radio emission. In Figure~\ref{fig:pulsedQ}, we also show the best fit line to just the periodically pulsing sources. For confirmed ECMI objects the brighter H$\alpha$ sources correspond to the brighter quiescent radio sources. The majority of the other quiescent radio emitters loosely cluster around this best fit line. The outliers, to the lower right, correspond preferentially to the warmer UCDs. This is likely a consequence of their H$\alpha$ emission having a significant chromospheric contribution. Consequently, the warmer brown dwarfs may be systematically farther to the right in Figure~\ref{fig:pulsedQ} than is warranted by any auroral contribution to their H$\alpha$ emission. Although it is unclear how this connection originates, especially considering the uncertain nature of the quiescent radio component, it is possible that both mechanisms rely not only on the same conditions that make a particular brown dwarf amenable to host the auroral electrodynamic engine, but both become stronger when that engine is more energetic, e.g., faster rotation, higher magnetospheric plasma densities, and stronger field strengths. Alternatively, the acceleration mechanism of the electrons responsible for the quiescent emission may be the same as that which accelerates the auroral electrons. A deeper understanding of the quiescent emission is needed to disentangle these physical effects.

\section{Conclusions}\label{sec:conc}

The observational shifts in stellar magnetic activity, into the ultracool dwarf regime, reflect the transition in physical properties in brown dwarfs going from stars to planets. The wide breadth of properties encompasses effects important in both the stellar and planetary cases. Our examination of the trends suggests that chromospheric and coronal heating begins to decay in very-low mass stars at the M/L transition, where the X-ray emission drops off dramatically. Relatively weaker heating continues into the L dwarfs, but is mostly suppressed in late L dwarfs and T dwarfs following the decline in atmospheric ionization fraction. If the stark drop off in X-rays at the M/L transition signals a steep decline in significant flare heating, it is possible that the residual chromospheres of L dwarfs may be sustained by a continued gradual decline in MHD wave dissipation with cooler atmospheric temperatures. A better understanding of the flare frequency distribution in the UCD regime will help elucidate the nature of this transition.

Amidst this transition some brown dwarf systems exhibit the conditions required to power an auroral electrodynamic engine. We stress that aurorae, as we have defined in this article, do not require an external star but can be generated internally by the brown dwarf system, as a consequence of large-scale magnetospheric currents. Although the exact conditions are unclear, they require strong magnetic field strengths, large-scale magnetic field topologies, fast rotation rates, and the presence of significant magnetospheric plasma. Although there are other possibilities (see Section~\ref{sec:aurorae}), the last condition is possibly associated with the presence of planets around brown dwarfs, in an analogy to the Jupiter-Io system, which seeds the magnetosphere with plasma through the moon's volcanic activity. Although the first three requirements are potentially met by most, if not all, brown dwarfs, the origins of the plasma and the role of a planetary satellite could be the underlying feature that distinguishes aurorally active brown dwarfs from inactive ones. This could potentially tie the auroral detection statistics to the planet formation rates around brown dwarfs and/or the presence of particular satellite system architectures.

The ECMI radio emission, periodically pulsed, coherent and with high degrees of circular polarization is the key observational indicator of the existence of the auroral magnetospheric processes. The aurorae are defined fundamentally by the presence of strong field aligned currents and a precipitating electron beam impacting the atmosphere. The consequences of this auroral beam include multi-wavelength surface emission features, like H$\alpha$ and H$_{2}$ Werner band emission in addition to the pulsed radio emission. Furthermore, the various multi-wavelength emissions provide significant probes of the brown dwarf atmospheres: temperature, ionization, chemistry. Moreover, these emissions may provide the only probes of brown dwarf upper atmospheres and could provide constraints on the properties of the energetic electron distribution and strength of the auroral currents. Because the atmospheric and electrodynamic conditions deviate considerably from the Jovian example, the expected energy balance of the different processes could be very different, however more observations are required to establish the breakdown of total auroral energy dissipation. Furthermore, auroral brown dwarfs provide an opportunity to explore a new parameter space in auroral physics, relative to what is seen in the Solar System: stronger magnetic field strengths, faster rotation rates, and warmer and denser atmospheres.

Understanding these effects properly will require further investigation in both observations of brown dwarfs and modeling the impact of these processes in their atmospheres. Much remains uncertain but, like the Sun and stellar activity, Jupiter and auroral activity will continue to provide important clues in deciphering the physics underlying brown dwarf magnetism at the cross section of stars and planets.

\section{Summary}\label{sec:summary}

In this article, we discussed how the trends in magnetic activity shift in the UCD regime and what the implications are for the underlying mechanisms powering magnetic phenomena in brown dwarf atmospheres, and in particular how auroral phenomena fit into these observations. Moreover, we applied the auroral paradigm to the multi-wavelength features of brown dwarf emission, illustrating the various processes and observational signatures indicative of UCD aurorae.

We summarize our main findings below:

\begin{itemize}
\item The distributions of H$\alpha$ emission in the UCD regime show a transition across L spectral types from predominantly coronal/chromospheric to likely auroral. 

\item The strength of X-ray and H$\alpha$ emissions in M7-M9 dwarfs are correlated over two orders of magnitude in X-ray luminosity. 

\item The predominant electrodynamic engine of auroral brown dwarfs may be co-rotation breakdown, and could involve the presence of close in planetary companions to the brown dwarfs. 

\item The auroral power is closely tied to the magnetic field strength and rotational velocity of the brown dwarf, regardless of the underlying engine. 

\item The presence of auroral emissions should be correlated with the presence of large-scale dipolar magnetic field topologies.

\item Brown dwarf auroral atmospheric emissions like H$\alpha$, H$_{3}^{+}$, and H$_{2}$ Werner band emission, are likely generated in auroral surface features but may be effected by atmospheric conditions. 

\item The auroral surface feature morphology is imprinted on the shape of the emission lines and depends on the electrodynamic engine powering the currents creating the auroral feature. 

\item Large amplitude NIR variability is likely dominated by transient cloud features and not magnetic effects, although the influence of auroral activity may be responsible for the long-lived sinusoidal features observed in photometric monitoring of some objects.

\item ECMI radio beaming patterns can produce a broad variety of observed light curves, accounting for the different morphologies of radio emissions from brown dwarfs hosting highly polarized periodic radio pulses. 

\item The observed radio variability can be strongly dependent on the relative geometry of the source and the observer. 

\item The potentially low degree of geometric selection effect in the observed detections of radio pulses suggest that brown dwarf magnetic axes may be significantly misaligned in general. 

\item Most quiescent radio sources with extended monitoring observations have also been detected as sources of periodic highly circularly polarized radio emission.

\item Quiescent radio luminosities are correlated with H$\alpha$ luminosities for confirmed periodically pulsing UCDs, suggesting a physical connection between the quiescent radio emission and the conditions generating brown dwarf auroral emission. 

\end{itemize}

\section*{Acknowledgments}

J.S.P. was supported by a grant from the National Science Foundation Graduate Research Fellowship under grant No. (DGE-11444469). J. S. P. would like to thank Jackie Villadsen for useful discussions in the development of arguments presented in this article. The authors would also like to thank the anonymous referee for a thorough reading of this manuscript and for providing critical feedback that greatly strengthened this work.

This research has benefitted from the M, L, T, and Y dwarf compendium housed at DwarfArchives.org. This research has benefitted from the Ultracool RIZzo Spectral Library maintained by Jonathan Gagn\'{e} and Kelle Cruz. This researched has benefitted from the Database of Ultracool Parallaxes maintained by Trent Dupuy.

This publication makes use of data products
from the Two Micron All Sky Survey, which is a
joint project of the University of Massachusetts
and the Infrared Processing and Analysis Cen-
ter/California Institute of Technology, funded by
the National Aeronautics and Space Administra-
tion and the National Science Foundation.

\section*{Appendix}

Throughout this paper, we have compiled the results of many literature sources measuring the different properties of UCDs. Below, we report the sources for the data used in this compilation, according to the measurement referenced. In many cases there have been multiple observations of the same properties for individual stars. In this article, we use the values which are more recent and illustrate consistency between the reports of multiple groups. 

\subsubsection*{Projected Rotational Velocities}

\citet{Mohanty2003,Mohanty2003b,Blake2010,Tanner2012,Prato2015,Gizis2016}; \citet[][and references therein]{Crossfield2014a}


\subsubsection*{H$\alpha$}

\citet{Kirkpatrick1999,Gizis2000,Kirkpatrick2000,Hall2002,Mohanty2003,Burgasser2003,Liebert2003,Reiners2007,Schmidt2007,Reiners2008,Lee2010,Burgasser2011,West2011,Gizis2013,Burgasser2015a,Metodieva2015,Pineda2016a}

\subsubsection*{Radio}

\citet{Berger2002,Berger2005,Burgasser2005a,Berger2006,Osten2006,Hallinan2007,Hallinan2008,Berger2009,Berger2010,McLean2012,Route2012, Antonova2013,Gizis2013,Route2013,Burgasser2015b,Osten2015,Williams2015a,Williams2015b,Route2016a, Route2016b,Kao2016,Lynch2016}

\subsubsection*{X-ray}

\citet{Tsuboi2003,Gizis2004,Berger2005,Audard2005,Berger2010,Osten2015}; \citet[][and references therein]{Cook2014,Williams2014}

\bibliographystyle{aasjournal}
\bibliography{panchromatic}

\end{document}